\documentclass[showpacs,aps,prl,twocolumn]{revtex4}
\usepackage{dcolumn}
\usepackage{graphicx}
\usepackage{amsmath}
\usepackage{latexsym}
\usepackage{amssymb}

\begin{document}
\title{A total measure of multi-particle quantum correlations in atomic Schr\"{o}dinger cat states}
\author{Ram Narayan Deb\\
Email: debram@rediffmail.com}
\address{Department of Physics, Chandernagore College,
 Chandernagore, Hooghly, Pin-712136, West Bengal, India}

\begin{abstract}
We propose a total measure of multi-particle quantum correlation in a system of $N$ two-level atoms (N qubits). We construct a parameter that encompasses all possible quantum correlations among $N$ two-level atoms in arbitrary symmetric pure states and define its numerical value to be the total measure of the net atom-atom correlations. We use that parameter to quantify the total quantum correlations in atomic 
Schr\"{o}dinger cat states, which are generated by the dispersive interaction in a cavity.  We study the variation of the net amount of quantum correlation as we vary the number of atoms from $N=2$ to $N=100$ and obtain some interesting results. We also study the variation of the net correlation, for fixed interaction time, as we increase the number of atoms in the excited state of the initial system, and notice some interesting features. We also observe the behaviour of the net quantum correlation as we continuously increase the interaction time, for the general state of $N$ two-level atoms in a dispersive cavity.  
\end{abstract}

\maketitle

\section{I. Introduction}
Over the past few years there has been a growing interest in studying quantum correlations in multiparticle systems\cite{Hald}-\cite{Ram2}. A lot of achievements have already been obtained in the study of bipartite quantum correlations \cite{Horodecki}. Among these one significant step is Wootters'formula \cite{Wootters} for the entanglement of formation for two-qubit mixed states and others include corresponding results for highly symmetrical states of higher dimensional systems \cite{Vollbrecht}, \cite{Terhal}. In Ref. \cite{Paternostro}, the sufficient and necessary conditions to induce entanglement on two remote qubits, by means of their respective linear interaction, with a two-mode driving field have been studied. A lot of work has also been done to understand the relationship between quantum entanglement and spin squeezing \cite{Sorensen}-\cite{Dalton}. In Ref. \cite{Giuseppe}, a complete set of generalised spin squeezing inequalities for detecting entanglement in an ensemble of qudits have been presented. It has also been shown, how to detect $k$-particle entanglement and bound entanglement. But, complete understanding of multi-particle quantum correlations is still an open and greater challenge.

In this paper, we propose a total measure of the multi-particle quantum correlations in systems of $N$ two-level atoms (N qubits) or equivalently in systems of $N$ spin-$\frac{1}{2}$ particles. We apply that measure to quantify the net amount of quantum correlations in atomic Sch\"{o}dinger cat states, which are correlated states, and play important role in understanding the conceptual foundations of quantum mechanics.

 In Ref. \cite{Ram1}, we proposed a partial measure of the multi-particle quantum correlations in a system of $N$ two-level atoms, because, that measure takes into account the quantum correlations of the atoms only in the $x$ and $y$-quadratures and does not take into account the correlations in the $z$-quadrature. In this paper, we show that, there are enough correlations also in  the $z$-quadrature, for a correlated state.
Therefore, to have a total measure of the atomic correlations, we must take into account the correlations along all the three mutually orthogonal quadratures, that is, along $x$, $y$, and $z$-quadratures.
 We, therefore, in this paper, construct the total measure of the multi-particle quantum correlations, by taking the root mean square value of all the atom-atom correlations along all the three mutually orthogonal directions.   
The parameter that we construct for the quantification of net correlation, named as quantum correlation parameter, encompasses all types of quantum correlations in a system of $N$ two-level atoms, and whose magnitude has been defined as a measure of the net amount of quantum correlations present in such systems. We use this method to analyze and quantify the amount of multi-particle quantum correlations in atomic Schr\"{o}dinger cat states.  Agarwal et al. \cite{Agarwal1} developed atomic Schr\"{o}dinger cat states using the dispersive interaction in a cavity. They applied time evolution to an initial atomic coherent state \cite{Arecchi} with the help of a Hamiltonian, non linear in atomic population inversion operator. At special times, the final states they obtained, are superpositions of several atomic coherent states each of different phases. The resultant states are 
Schr\"{o}dinger cat states which show interesting quantum interferences. Furthermore, the resultant states are highly correlated states, and this is our motivation to quantify the amount of quantum correlations present in such system. We study the variation of the amount of quantum correlations as we increase the number of atoms in the system from $N=2$ to $N=100$. We notice some interesting features. We also analyse the situation as we continuously increase the number of atoms in the excited state of the initial system. We observe that, when the number of atoms in the assembly is an odd number, the net quantum 
correlation behaves in a different way as compared to the case, when the total number of atoms is an even number. We also study the variation of the net correlation with the interaction time, for a general state of two-level atoms in a dispersive cavity. 

The organization of the paper is as follows. In section II, we construct the total measure of multi-particle quantum correlation, that is used to quantify the net amount of quantum correlation in a system of $N$ two-level atoms. In section III, we quantify the amount of correlation in atomic 
Schr\"{o}dinger cat states. In section IV, we present the summary and conclusion.

\section{II. Construction of a total measure of quantum correlation}

At first, we mention here that, in Ref. \cite{Ram1} and \cite{Ram2}, we have developed a partial measure of multi-particle quantum correlations in a system of $N$ two-level atoms, because, in those papers we considered only the quantum correlations along the $x$ and $y$-quadratures only. But, in this paper, we show that there are lot of correlations in the $z$-quadrature also. Therefore, to construct a total measure of multi-particle quantum correlations, we must take into account all the correlations in all the three mutually orthogonal quadratures, that is, in the $x$, $y$, and $z$-quadratures. So, in this paper, we develop a total measure of multi-particle quantum correlations, by taking into account all the quantum correlations in all the three mutually orthogonal quadratures, that is, in $x$, $y$, and $z$-quadratures, and
apply that total measure to quantify the net amount of quantum correlations in atomic Schr\"{o}dinger cat states.  Now, we, mention here that, in this paper, in our way toward the construction of a total measure of quantum correlations, the method of extracting out the quantum correlations in the $x$ and $y$-quadratures have already been developed by us in Ref. \cite{Ram1} and \cite{Ram2}. But, in this paper, we provide a complete method for the construction of a total measure of multi-particle quantum correlations.

Now, in order to construct a total measure of multi-particle quantum correlations
for a system of $N$ two-level atoms ($N$ qubits), our aim is to first extract out all possible quantum correlations that exist among all the atoms in the system. The quantum correlations that exist among the atoms are expressed here mathematically in terms of the expectation values of the pseudo-spin operators over the composite state of all the two-level atoms. We, therefore, first discuss about the pseudo-spin operator of a two-level atom (a single qubit).

If the upper and lower energy level of the $n$-th atom in the assembly  
are denoted as $|u_n\rangle$ and $|l_n\rangle$ respectively, then,  we can construct some operators, (with $\hbar = 1$) as \cite{Wineland1}
\begin{eqnarray}
\hat{J}_{n_x} &=& (1/2)\big(|u_n\rangle\langle l_n| + |l_n\rangle
\langle u_n|\big),\label{1.1a1}\\
\hat{J}_{n_y} &=& (-i/2)\big(|u_n\rangle\langle l_n| - |l_n\rangle
\langle u_n|\big),\label{1.1a2}\\
\hat{J}_{n_z} &=& 
(1/2)\big(|u_n\rangle\langle u_n| - |l_n\rangle\langle l_n|
\big).
\label{1.1a3}
\end{eqnarray}
We can check that these operators satisfy
\begin{eqnarray}
[\hat{J}_{n_x}, \hat{J}_{n_y}] &=& i\hat{J}_{n_z},~~~~ [\hat{J}_{n_y}, \hat{J}_{n_z}] = i\hat{J}_{n_x},\nonumber\\
&& [\hat{J}_{n_z}, \hat{J}_{n_x}] = i\hat{J}_{n_y},
\label{1.2}
\end{eqnarray} 
which are exactly the similar commutation relations as those satisfied by the Pauli spin operators. For this reason the above operators are termed as pseudo-spin operators.

Now, for a system of $N$ such two-level atoms, we construct collective pseudo-spin operators,
\begin{eqnarray}
\hat{J}_x = \sum_{i=1}^{N}\hat{J}_{i_{x}},~~~~ 
\hat{J}_y = \sum_{i=1}^{N}\hat{J}_{i_{y}},~~~~ 
\hat{J}_z = \sum_{i=1}^{N}\hat{J}_{i_{z}}.
\label{1.3}
\end{eqnarray}
Now, since the individual atomic operators satisfy
\begin{eqnarray}
\big[\hat{J}_{1_x}, \hat{J}_{2_z}\big] = 0,~
\big[\hat{J}_{1_x}, \hat{J}_{1_y}\big] = i\hat{J}_{1_z}, ~
\big[\hat{J}_{2_y}, \hat{J}_{2_z}\big] = i\hat{J}_{2_x},
\label{1.2a1}
\end{eqnarray}
and so on,
the collective pseudo-spin operators satisfy,
\begin{eqnarray}
[\hat{J}_x , \hat{J}_y] &=& i\hat{J}_z,~~~~[\hat{J}_y , \hat{J}_z] = i\hat{J}_x,\nonumber\\
&& [\hat{J}_z , \hat{J}_x] = i\hat{J}_y.
\label{1.2a2}
\end{eqnarray} 
Now, the collective quantum state vector for $N$ two-level atoms is represented as $|\psi_j\rangle$, where the quantum number $j$ is related to the number of atoms $N$ as $j = N/2$. We can express this composite state vector $|\psi_j\rangle$ as a linear superposition of the basis states $|j,m\rangle$, where $|j,m\rangle$ are the simultaneous eigenvectors of the operators  $\hat{J}^2 = \hat{J}_x^2
+ \hat{J}_y^2 + \hat{J}_z^2$ and $\hat{J}_z$, with eigenvalues $j(j+1)$ and 
$m$ respectively.
The quantum number $m$ is related to $j$ as $m = -j, -j+1, ....(j-1), j$.

 We define the mean pseudo-spin vector as
\begin{equation}
\langle\hat{\mathbf{J}}\rangle = \langle\hat{J}_x\rangle
\hat{i} + \langle\hat{J}_y\rangle\hat{j} + \langle\hat{J}_z\rangle\hat{k},\label{1.4}
\end{equation}
where the average values in the above expression are to be calculated over the state $|\psi_j\rangle$ and
$\hat{i}$, $\hat{j}$ and $\hat{k}$ are the unit vectors along positive $x$, $y$ and $z$ axes respectively.
Now, the quantum fluctuations in $\hat{J}_x$, $\hat{J}_y$, and $\hat{J}_z$ for the state $|\psi_j\rangle$, are defined as
\begin{equation}
\Delta J_{x, y, z} =\sqrt{\langle\psi_j|{\hat{J}_{x, y, z}}^2|\psi_j\rangle - \langle\psi_j|\hat{J}_{x, y, z}|\psi_j\rangle^2}.
\label{2.5}
\end{equation}
Now, the state $|\psi_j\rangle$ is said to be a coherent spin state (CSS) \cite{Kitagawa},
or an atomic coherent state in literature, if in a plane, normal to 
$\langle\hat{\mathbf{J}}\rangle$, the quantum fluctuations of the pseudo-spin operators along any two mutually orthogonal directions are same, and have value $\sqrt{N}/2$, that is, 

\begin{equation}
\Delta J_{1}^2 = \Delta J_{2}^2 = \frac{j}{2} = \frac{N}{4},
\label{2.5a}
\end{equation}
where $\hat{J}_1$ and $\hat{J}_2$ are the pseudo-spin operators along any two mutually orthogonal directions, in a plane perpendicular to $\langle\hat{\mathbf{J}}\rangle$.

The state $|\psi_j\rangle$ is said to be a squezeed spin state (SSS) 
 \cite{Kitagawa}, if
\begin{equation}
\Delta J_{1}^2 ~~~ or,~~~ \Delta J_{2}^2 <  \frac{j}{2} = \frac{N}{4}.
\label{2.5b}
\end{equation}

Now, coventionally we rotate our coordinate system $\{x,y,z\}$ to 
$\{x^\prime,y^\prime,z^\prime\}$, so that the mean pseudo-spin vector
$\langle\hat{\mathbf{J}}\rangle$ points along the $z^\prime$ axis, and we take the pseudo-spin operators $\hat{J}_1$ and $\hat{J}_2$ as $\hat{J}_{x^\prime}$ and $\hat{J}_{y^\prime}$ respectively. In that case, we have for a coherent spin state
$\Delta J_{x^\prime}^2 = \Delta J_{y^\prime}^2 = j/2 = N/4$, and for a squeezed spin state we have $\Delta J_{x^\prime}^2$ or, $\Delta J_{y^\prime}^2 < j/2 = N/4$.

The need to investigate the square of the quantum fluctuations, 
$\Delta J_1^2$ and $\Delta J_2^2$ or equivalently $\Delta J_{x^\prime}^2$ and $\Delta J_{y^\prime}^2$, in a plane normal to 
$\langle\hat{\mathbf{J}}\rangle$, for defining a coherent spin state and a squeezed spin state, is to exclude the mere mathematical coordinate dependency and include quantum correlations in the notion of squeezing 
\cite{Kitagawa}. If we do not consider the quantum fluctuations in mutually orthogonal directions in a plane normal to $\langle\hat{\mathbf{J}}\rangle$, then an atomic coherent state appears to be a squeezed state \cite{Ram3}, which is misleading. Therefore, to find out wheteher an arbitrary atomic state is a coherent state or a squeezed state, we must investigate the quantum fluctuations in mutually orthogonal directions, in a plane normal to $\langle\hat{\mathbf{J}}\rangle$.

We now show that the square of the quantum fluctuations $\Delta J_{x^\prime}^2$, $\Delta J_{y^\prime}^2$, and $\Delta J_{z^\prime}^2$ can be expressed as algebraic sum of the square of the corresponding quantum fluctuations of $N$ individual atoms and all possible combinations of correlations among the $N$ atoms.
Our ability to write the square of the quantum fluctuations in the above mentioned way gives us the opportunity to extract out only the all possible atom-atom correlation terms and construct the total measure of quantum correlation for such systems.

Now, an arbitrary symmetric pure state for a system of $N$ two-level atoms in the
$\{ m_1, m_2, m_3,....m_N \}$ representation is given as

\begin{eqnarray}
|\Psi\rangle &=& A_1\bigg\vert\frac{1}{2},\frac{1}{2},....\frac{1}{2}
\bigg\rangle
+ \frac{A_2}{\sqrt{{}^NC_1}}\Bigg[\bigg\vert-\frac{1}{2},\frac{1}{2},\frac{1}{2},....\frac{1}{2}
\bigg\rangle \nonumber\\
&+& \bigg\vert\frac{1}{2},-\frac{1}{2},\frac{1}{2},....\frac{1}{2}
\bigg\rangle + ....\bigg\vert\frac{1}{2},\frac{1}{2},\frac{1}{2},....
-\frac{1}{2}\bigg\rangle\Bigg]\nonumber\\ 
&+& \frac{A_3}{\sqrt{{}^NC_2}}\Bigg[\bigg\vert-\frac{1}{2},-\frac{1}{2},\frac{1}{2},....\frac{1}{2}
\bigg\rangle \nonumber\\
&+& \bigg\vert-\frac{1}{2},\frac{1}{2},-\frac{1}{2},....\frac{1}{2}
\bigg\rangle + .... \bigg\vert\frac{1}{2},\frac{1}{2},....
-\frac{1}{2},-\frac{1}{2}\bigg\rangle\Bigg]\nonumber\\
&+& ............A_{N+1}\bigg\vert-\frac{1}{2},-\frac{1}{2},-
\frac{1}{2},....-\frac{1}{2}\bigg\rangle,
\label{2.5c}
\end{eqnarray}

where $A_1$, $A_2$, ..., $A_{N+1}$ are constants and
 ${}^N C_r$ is given as
\begin{equation}
{}^NC_r = \frac{N!}{r!(N-r)!}.
\label{2.5d}
\end{equation}

Now, the mean pseudo-spin vector $\langle\hat{\mathbf{J}}\rangle$ for the above state points in an arbitrary direction in space.
To investigate the quantum fluctuations $\Delta J_{x^\prime}^2$ and
$\Delta J_{y^\prime}^2$ in a plane normal to 
$\langle\hat{\mathbf{J}}\rangle$, we perform a rotation of the coordinate system from $\{x, y, z\}$ to $\{x^\prime, y^\prime, z^\prime\}$, in such a way that $\langle\hat{\mathbf{J}}\rangle$ points along the $z^\prime$ axis.
As a result we obtain 
\begin{eqnarray}
\hat{J}_{x^\prime} &=& \hat{J}_x\cos\theta\cos\phi + \hat{J}_y
\cos\theta\sin\phi - \hat{J}_z\sin\theta\label{2.5d1}\\
\hat{J}_{y^\prime} &=& -\hat{J}_x\sin\phi + \hat{J}_y\cos\phi
\label{2.5d2}\\
\hat{J}_{z^\prime} &=& \hat{J}_x\sin\theta\cos\phi + \hat{J}_y
\sin\theta\sin\phi + \hat{J}_z\cos\theta,\label{2.5d3}
\end{eqnarray}
where,
\begin{eqnarray}
\cos\theta &=& \frac{\langle\hat{J}_z\rangle}
{|\langle\hat{\mathbf{J}}\rangle|}\label{2.5d4}\\
\cos\phi &=& \frac{\langle\hat{J}_x\rangle}{\sqrt{\langle\hat{J}_x\rangle^2 + \langle\hat{J}_y\rangle^2}}.
\label{2.5d5}
\end{eqnarray}
We can observe from Eqs. (\ref{2.5d1}), (\ref{2.5d2}), 
(\ref{2.5d4}), and (\ref{2.5d5}), that $\langle\hat{J}_{x^\prime}\rangle = 0$, $\langle\hat{J}_{y^\prime}\rangle = 0$, and consequently the mean pseudo-spin vector points along the $z^\prime$ axis.

Now, using Eqs. (\ref{2.5d1}) and (\ref{2.5d2}), we observe that the square of the quantum fluctuations $\Delta{J_{x^\prime}^2}$ and $\Delta{J_{y^\prime}^2}$ have the form
\begin{eqnarray}
&&\Delta{J_{x^\prime}^2} = \langle J_{x^\prime}^2 \rangle
- \langle J_{x^\prime} \rangle^2\nonumber\\ 
&=&\Delta{J_x}^2\cos^2\theta\cos^2\phi + \Delta{J_y}^2\cos^2\theta\sin^2\phi\nonumber \\
&+& \Delta{J_z}^2\sin^2\theta + \Big{(}\langle{J_x}{J_y} + J_y J_x\rangle - 2\langle J_x\rangle\langle J_y\rangle\Big{)}\nonumber\\
&\times&\cos^2\theta\sin\phi\cos\phi - \Big{(}\langle{J_x}{J_z} + J_z J_x\rangle - 2\langle J_x\rangle\langle J_z\rangle\Big{)}\nonumber\\
&\times&\sin\theta\cos\theta\cos\phi - \Big{(}\langle{J_y}{J_z} + J_z J_y\rangle - 2\langle J_y\rangle\langle J_z\rangle\Big{)}\nonumber\\
&\times&\sin\theta\cos\theta\sin\phi
\label{2.5d6}
\end{eqnarray}
and
\begin{eqnarray}
\Delta{J_{y^\prime}^2} &=& \langle J_{y^\prime}^2 \rangle
- \langle J_{y^\prime} \rangle^2\nonumber\\ 
&=&\Delta {J_x}^2\sin^2\phi + \Delta {J_y}^2\cos^2\phi - \Big{(}\langle{J_x}{J_y} + J_y J_x\rangle\nonumber\\
 &-& 2\langle J_x\rangle\langle J_y\rangle\Big{)}\sin\phi\cos\phi.
\label{2.5d7}
\end{eqnarray}
Now using Eqs. (\ref{1.3}) and (\ref{2.5d6}), we can express 
$\Delta{J_{x^\prime}^2}$ as
\begin{eqnarray}
&&\Delta J_{x^\prime}^2 = \sum_{i=1}^{N}\Bigg{[}\Delta
J_{i_{x}}^2\cos^2\theta\cos^2\phi + \Delta J_{i_{y}}^2\cos^2\theta\sin^2\phi  \nonumber\\
&+& \Delta J_{i_{z}}^2\sin^2\theta +\Big{(}\langle
\hat{J}_{i_{x}}\hat{J}_{i_{y}}+\hat{J}_{i_{y}}\hat{J}_
{i_{x}}\rangle - 2\langle\hat{J}_{i_{x}}\rangle\langle
\hat{J}_{i_{y}}\rangle\Big{)}\nonumber\\
&\times&\cos^2\theta\sin\phi\cos\phi - \Big{(}\langle
\hat{J}_{i_{x}}\hat{J}_{i_{z}}+\hat{J}_{i_{z}}\hat{J}_
{i_{x}}\rangle - 2\langle\hat{J}_{i_{x}}\rangle\langle
\hat{J}_{i_{z}}\rangle\Big{)}\nonumber\\ 
&\times&\sin\theta\cos\theta\cos\phi - \Big{(}\langle
\hat{J}_{i_{y}}\hat{J}_{i_{z}}+\hat{J}_{i_{z}}\hat{J}_
{i_{y}}\rangle - 2\langle\hat{J}_{i_{y}}\rangle\langle
\hat{J}_{i_{z}}\rangle\Big{)}\nonumber\\
&\times&\sin\theta\cos\theta\sin\phi \Bigg{]} + 
\sum_{i=1}^{N} \sum_{{}^{l=1}_{l\ne i}}^{N}\Bigg{[}\Big{(}\langle\hat{J}_
{i_{x}}\hat{J}_{l_{x}}\rangle - \langle\hat{J}_
{i_{x}}\rangle\langle\hat{J}_{l_{x}}\rangle\Big{)}\nonumber\\
&\times&\cos^2\theta\cos^2\phi + \Big{(}\langle\hat{J}_
{i_{y}}\hat{J}_{l_{y}}\rangle - \langle\hat{J}_
{i_{y}}\rangle\langle\hat{J}_{l_{y}}\rangle\Big{)}\cos^2\theta\sin^2\phi\nonumber\\
&+& \Big{(}\langle\hat{J}_
{i_{z}}\hat{J}_{l_{z}}\rangle - \langle\hat{J}_
{i_{z}}\rangle\langle\hat{J}_{l_{z}}\rangle\Big{)}\sin^2\theta + 2\Big{(}\langle\hat{J}_
{i_{x}}\hat{J}_{l_{y}}\rangle\nonumber\\
&-& \langle\hat{J}_
{i_{x}}\rangle\langle\hat{J}_{l_{y}}\rangle\Big{)}\cos^2\theta\sin\phi\cos\phi - 2\Big{(}\langle\hat{J}_
{i_{x}}\hat{J}_{l_{z}}\rangle\nonumber\\
 &-& \langle\hat{J}_
{i_{x}}\rangle\langle\hat{J}_{l_{z}}\rangle\Big{)}\sin\theta\cos\theta\cos\phi - 2\Big{(}\langle\hat{J}_
{i_{y}}\hat{J}_{l_{z}}\rangle - \langle\hat{J}_
{i_{y}}\rangle\langle\hat{J}_{l_{z}}\rangle\Big{)}\nonumber\\
&\times& \sin\theta\cos\theta\sin\phi\Bigg{]}.
\label{2.5d18}
\end{eqnarray}
Now, comparing the term under the single summation symbol in the above equation with Eq. (\ref{2.5d6}), we find that it is nothing but
$\Delta J_{i_{x^\prime}}^2$, that is, the square of the quantum fluctuation in the $x^\prime$-quadrature of the $i$-th atom in the assembly. Now, calling the 
term with the double summation symbol in the above equation as 
$C_X$, we can write Eq. (\ref{2.5d18}) as,
\begin{equation}
\Delta J_{x^\prime}^2 = \sum_{i=1}^{N}\Delta J_{i_{x^\prime}}^2 + C_X.
\label{2.5d19}
\end{equation} 
We observe that $C_X$ is solely made up of all possible bipartite quantum correlations among the $N$ two-level atoms.   
Thus, we have expressed the square of the quantum fluctuation $\Delta J_{x^\prime}^2$ for the composite state of $N$ two-level atoms as an algebraic sum of the square of the quantum fluctuations 
$\Delta J_{i_{x^\prime}}^2$ of the $N$ individual atoms 
and all possible bipartite quantum correlations, $C_X$, among the $N$ atoms.

We now present a similar expression for 
$\Delta J_{y^\prime}^2$. Using Eqs. (\ref{1.3}) and (\ref{2.5d7}), we can
write
\begin{eqnarray}
\Delta J_{y^\prime}^2 &=& \sum_{i=1}^{N}\Bigg{[}\Delta
J_{i_{x}}^2\sin^2\phi + \Delta J_{i_{y}}^2\cos^2\phi  \nonumber\\
&-& \Big{(}\langle
\hat{J}_{i_{x}}\hat{J}_{i_{y}}+\hat{J}_{i_{y}}\hat{J}_
{i_{x}}\rangle - 2\langle\hat{J}_{i_{x}}\rangle\langle
\hat{J}_{i_{y}}\rangle\Big{)}
\sin\phi\cos\phi\nonumber\Bigg{]}\nonumber\\
&+& \sum_{i=1}^{N} \sum_{{}^{l=1}_{l\ne i}}^{N}\Bigg{[}\Big{(}\langle
\hat{J}_{i_{x}}\hat{J}_{l_{x}}\rangle - \langle\hat{J}_
{i_{x}}\rangle\langle\hat{J}_{l_{x}}\rangle\Big{)}
\sin^2\phi\nonumber\\
&+& \Big{(}\langle\hat{J}_
{i_{y}}\hat{J}_{l_{y}}\rangle - \langle\hat{J}_
{i_{y}}\rangle\langle\hat{J}_{l_{y}}\rangle\Big{)}
\cos^2\phi
+ 2\Big{(}\langle\hat{J}_
{i_{x}}\hat{J}_{l_{y}}\rangle\nonumber\\
&-& \langle\hat{J}_
{i_{x}}\rangle\langle\hat{J}_{l_{y}}\rangle\Big{)}\sin\phi\cos\phi\Bigg{]}.
\label{2.5d21}
\end{eqnarray}
 We observe that the term under the single summation symbol in the above equation is just like the right hand side of Eq. (\ref{2.5d7}), and hence it is nothing but 
$\Delta J_{i_{y^\prime}}^2$, that is, the square of the quantum fluctuation in the $y^\prime$-quadrature of the $i$-th atom in the assembly. 
Now, calling the term with the double summation symbol in 
Eq. (\ref{2.5d21}) as $C_Y$, we can write the above equation as 
\begin{equation}
\Delta J_{y^\prime}^2 = \sum_{i=1}^{N}\Delta J_{i_{y^\prime}}^2 + C_Y.
\label{2.5d22}
\end{equation}
 We observe that 
$C_Y$ is made up of all bipartite quantum correlations among the 
$N$ two-level atoms. Thus, we have expressed the square of the quantum fluctuation in $\hat{J}_{y^\prime}$ of the entire system as an algebraic sum of the square of the corresponding quantum fluctuations of the $N$ individual atoms, and all possible quantum correlations among them.

Using similar analysis, we can write the square of the quantum fluctuation in $\hat{J}_{z^\prime}$, that is, $\Delta J_{z^\prime}^2$ for the composite state of $N$ two-level atoms as
\begin{eqnarray}
\Delta J_{z^\prime}^2 &=& \sum_{i=1}^{N}\Bigg{[}\Delta J_{i_{x}}^2\cos^2\theta\sin^2\phi + \Delta J_{i_{y}}^2\sin^2\theta\sin^2\phi\nonumber\\
&+& 
\Delta J_{i_{z}}^2\cos^2\phi + \Big{(}\langle\hat{J}_{i_{x}}
\hat{J}_{i_{y}} + \hat{J}_{i_{y}}\hat{J}_{i_{x}}\rangle - 
2\langle\hat{J}_{i_{x}}\rangle\langle\hat{J}_{i_{y}}\rangle\Big{)}
\nonumber\\
&\times&\sin\theta\cos\theta\sin^2\phi+\Big{(}\langle\hat{J}_{i_{x}}
\hat{J}_{i_{z}} + \hat{J}_{i_{z}}\hat{J}_{i_{x}}\rangle\nonumber\\ 
&-& 
2\langle\hat{J}_{i_{x}}\rangle\langle\hat{J}_{i_{z}}\rangle\Big{)}\cos\theta\sin\phi\cos\phi + \Big{(}\langle\hat{J}_{i_{y}}
\hat{J}_{i_{z}}\nonumber\\ 
&+& \hat{J}_{i_{z}}\hat{J}_{i_{y}}\rangle
- 2\langle\hat{J}_{i_{y}}\rangle\langle\hat{J}_{i_{z}}\rangle\Big{)}
\sin\theta\sin\phi\cos\phi\Bigg{]}  \nonumber\\
&+&\sum_{i=1}^{N}\sum_{{}^{l=1}_{l\ne i}}^{N}\Bigg{[}\Big{(}\langle\hat{J}_{i_{x}}\hat{J}_{l_{x}}\rangle - \langle\hat{J}_{i_{x}}\rangle\langle
\hat{J}_{l_{x}}\rangle\Big{)}\cos^2\theta\sin^2\phi\nonumber\\
&+& \Big{(}\langle\hat{J}_{i_{y}}\hat{J}_{l_{y}}\rangle - \langle\hat{J}_{i_{y}}\rangle\langle
\hat{J}_{l_{y}}\rangle\Big{)}\sin^2\theta\sin^2\phi\nonumber\\
&+&\Big{(}\langle\hat{J}_{i_{z}}\hat{J}_{l_{z}}\rangle - \langle\hat{J}_{i_{z}}\rangle\langle
\hat{J}_{l_{z}}\rangle\Big{)}\cos^2\phi\nonumber\\
&+&2\Big{(}\langle\hat{J}_{i_{x}}\hat{J}_{l_{y}}\rangle - \langle\hat{J}_{i_{x}}\rangle\langle
\hat{J}_{l_{y}}\rangle\Big{)}\sin\theta\cos\theta\sin^2\phi\nonumber\\
&+&2\Big{(}\langle\hat{J}_{i_{x}}\hat{J}_{l_{z}}\rangle - \langle\hat{J}_{i_{x}}\rangle\langle
\hat{J}_{l_{z}}\rangle\Big{)}\cos\theta\sin\phi\cos\phi\nonumber\\
&+&2\Big{(}\langle\hat{J}_{i_{y}}\hat{J}_{l_{z}}\rangle - \langle\hat{J}_{i_{y}}\rangle\langle
\hat{J}_{l_{z}}\rangle\Big{)}\sin\theta\sin\phi\cos\phi\Bigg{]}.
\label{2.5d22a1}
\end{eqnarray}
The term under the single summation symbol in the above equation is actually 
$\Delta J_{i_{z^\prime}}^2$,that is the quantum fluctuation in the $z^\prime$-quadrature of the $i$-th atom in the assembly and, the term under the double summation symbol
represent the atom-atom correlations. We notice from the above equation that, there is full of atom-atom correlations along the $z^\prime$-direction. Therefore, to have a total measure of the atomic correlations, we must take into account of the above correlations, expressed in the above equation.

  We, thus, have expressed the square of the quantum fluctuation in the $z^\prime$-quadrature of the composite system as a sum of the square of the  quantum fluctuations, $\Delta J_{i_{z^\prime}}^2$, of all the individual atoms and the correlation terms among them. Therefore, we can write the above equation as
\begin{equation}
\Delta J_{z^\prime}^2 = \sum_{i=1}^{N}\Delta J_{i_{z^\prime}}^2 + C_Z,
\label{2.5d22a2}
\end{equation}
where $C_Z$ is the term with the double summation in Eq. 
(\ref{2.5d22a1}) and represents the atomic correlations in the 
$z^\prime$-quadrature.

Now, if atoms in the assembly are uncorrelated, then, the composite quantum state vector $|\Psi\rangle$ can be expressed as the product of the individual atomic state vectors, that is
\begin{equation}
|\Psi\rangle = |\Psi_1\rangle \otimes |\Psi_2\rangle \otimes |\Psi_3\rangle......\otimes |\Psi_N\rangle,
\label{2.5d23}
\end{equation}
where $|\Psi_1\rangle$, $|\Psi_2\rangle$, ...
$|\Psi_N\rangle$ are the individual atomic state vectors.
In that case, we have for $i \ne l$,
\begin{eqnarray}
\langle\hat{J}_{i_{x}}\hat{J}_{l_{x}}\rangle &=& \langle
\hat{J}_{i_{x}}\rangle\langle\hat{J}_{l_{x}}\rangle,~~~~
\langle\hat{J}_{i_{y}}\hat{J}_{l_{y}}\rangle = \langle
\hat{J}_{i_{y}}\rangle\langle\hat{J}_{l_{y}}
\rangle,\nonumber\\
\langle\hat{J}_{i_{z}}\hat{J}_{l_{z}}\rangle &=& \langle
\hat{J}_{i_{z}}\rangle\langle\hat{J}_{l_{z}}\rangle,~~~~
\langle\hat{J}_{i_{x}}\hat{J}_{l_{y}}\rangle = \langle
\hat{J}_{i_{x}}\rangle\langle\hat{J}_{l_{y}}\rangle,
\nonumber\\
\langle\hat{J}_{i_{x}}\hat{J}_{l_{z}}\rangle &=& \langle
\hat{J}_{i_{x}}\rangle\langle\hat{J}_{l_{z}}\rangle,~~~~
\langle\hat{J}_{i_{y}}\hat{J}_{l_{z}}\rangle = \langle
\hat{J}_{i_{y}}\rangle\langle\hat{J}_{l_{z}}\rangle.
\label{2.5d24}
\end{eqnarray}
If these conditions are satisfied then,
all the terms under the double summation symbols in Eqs. (\ref{2.5d18}), (\ref{2.5d21}), and (\ref{2.5d22a1}) are zero. Therefore, when the atoms in the assembly are uncorrelated, we have
\begin{equation}
C_X = C_Y = C_Z = 0.
\label{2.5d25}
\end{equation}

Now, we can calculate and find that the square of the quantum fluctuations 
in $x^\prime$ and $y^\prime$-quadratures of the individual constituent atoms in the assembly have value $1/4$, that is

\begin{eqnarray}
\Delta J_{1_{x^\prime}}^2 &=& \Delta J_{2_{x^\prime}}^2
=...\Delta J_{N_{x^\prime}}^2 = \frac{1}{4},
\label{2.8}\\
\Delta J_{1_{y^\prime}}^2 &=& \Delta J_{2_{y^\prime}}^2
=...\Delta J_{N_{y^\prime}}^2 = \frac{1}{4}.
\label{2.9}
\end{eqnarray} 
Therefore, using Eqs. (\ref{2.8}) and (\ref{2.9}), we can write 
Eqs. (\ref{2.5d19}), and (\ref{2.5d22}) respectively, as 
\begin{eqnarray}
\Delta J_{x^\prime}^2  &=& \frac{N}{4} + C_X,
\label{2.10a1}\\
\Delta J_{y^\prime}^2 &=&  \frac{N}{4} + C_Y.
\label{2.10b1}
\end{eqnarray}

Now, the quantum fluctuation $\Delta J_{i_{z^\prime}}^2$ is
\begin{equation}
\Delta J_{i_{z^\prime}}^2 = \frac{1}{4} - 
|\langle{\bf \hat{J}_i}\rangle|^2,
\label{2.9a1}
\end{equation}
where $\langle{\bf \hat{J}_i}\rangle$ is the mean pseudo-spin vector of the $i$-th atom in the assembly.

Now, since the composite state of $N$ atoms in Eq. (\ref{2.5c}) is a symmetric state, where all the atoms have been treated on equal footing, the mean pseudo-spin vector $\langle{\bf \hat{J}}\rangle$ of the composite system is related with the mean pseudo-spin vector 
$\langle{\bf \hat{J}_i}\rangle$ of the individual atoms as
\begin{eqnarray}
\langle{\bf \hat{J}}\rangle &=& \langle\hat{J}_x\rangle \hat{i} + \langle\hat{J}_y\rangle \hat{j} + \langle\hat{J}_z\rangle \hat{k}\nonumber\\
&=& N \langle\hat{J}_{i_{x}}\rangle\hat{i} + N \langle\hat{J}_
{i_{y}}\rangle\hat{j} + N\langle\hat{J}_{i_{z}}\rangle \hat{k}\nonumber\\
&=& N \langle{\bf \hat{J}_i}\rangle.
\label{2.9a2}
\end{eqnarray}
Therefore, using Eq. (\ref{2.9a2}) in Eq. (\ref{2.9a1}), we obtain
\begin{equation}
\Delta J_{i_{z^\prime}}^2 = \frac{1}{4} - \frac{1}{N^2}|\langle{\bf \hat{J}}\rangle|^2.
\label{2.9a3}
\end{equation}

And, finaly using Eq. (\ref{2.9a3}) in Eq. (\ref{2.5d22a2}), we get
\begin{eqnarray}
\Delta J_{z^\prime}^2 &=& \sum_{i=1}^{N}\Big{(}\frac{1}{4} - \frac{1}{N^2}|\langle{\bf \hat{J}}\rangle|^2\Big{)} + C_Z\nonumber\\
&=& \frac{N}{4} - \frac{1}{N}|\langle{\bf \hat{J}}\rangle|^2 + C_Z
\label{2.9a4}
\end{eqnarray}

Now, for a coherent spin state, which is a completely separable state, the square of the quantum fluctuations in $x^\prime$ and $y^\prime$-quadratures, that is, $\Delta J_{x^\prime}^2$ and $\Delta J_{y^\prime}^2$
have value $N/4$ \cite{Kitagawa}. 
Therefore, we can write Eqs. (\ref{2.10a1}) and (\ref{2.10b1}) as
\begin{eqnarray}
\Delta J_{x^\prime}^2 - \Delta J_{x^\prime}^2|_{un-corr.} &=& C_X~,\label{2.10a}\\
\Delta J_{y^\prime}^2 - \Delta J_{y^\prime}^2|_{un-corr.} &=& C_Y.
\label{2.10b}
\end{eqnarray}

Therefore, $C_X$ and 
$C_Y$ are equal to the deviations of the square of the quantum fluctuations in $\hat{J}_{x^\prime}$ and 
$\hat{J}_{y^\prime}$ respectively, of the arbitrary state 
$|\Psi\rangle$ from those of an uncorrelated state. We also notice from Eqs. (\ref{2.10a1}), and (\ref{2.10b1}), that, if $C_X < 0$, we have spin 
squeezing in the $x^\prime$-quadrature of the system, and in that case 
$C_Y$ should be greater than zero, and vice-versa.

Now, regarding the quantum fluctuation in the $z^\prime$-quadrature, it is to be mentioned here that $\Delta J_{z^\prime}^2 = 0$ for a completely seperable state. In that case, all the conditions in Eqs. 
(\ref{2.5d24}) are satisfied, producing $C_Z =0$, and $|\langle{\bf \hat{J}}\rangle|^2$ attains the value $N^2/4$, thus, making 
$\Delta J_{z^\prime}^2 = 0$ in Eq. (\ref{2.9a4}).  

We notice from Eqs. (\ref{2.5d18}), (\ref{2.5d19}), 
(\ref{2.5d21}), (\ref{2.5d22}), (\ref{2.5d22a1}), and (\ref{2.5d22a2}) that $C_X$, 
$C_Y$, and $C_Z$ contain only the all possible quantum correlations among the atoms, and they are the quantum correlation terms in all the three mutually orthogonal directions, that is, the $x^\prime$, $y^\prime$, and
$z^\prime$-directions. Therefore,
we, use these correlation terms to construct a total measure of multi-particle quantum correlations in a system of $N$ two-level atoms (N qubits). Now, since, 
$C_X$, $C_Y$, and $C_Z$ may be positive or negative, we cannot simply add them to calculate the net quantum correlation among the atoms. We, therefore, take the root mean square value of the above three correlation terms to construct the total measure of multi-particle quantum correlations. So, we define the net quantum correlation parameter as
\begin{equation}
S = \frac{1}{\sqrt{3}}\Big{[}(C_X)^2 + (C_Y)^2 + (C_Z)^2\Big{]}^{1/2}.
\label{2.11}
\end{equation}   
Thus, $S$ is the root mean squared value of all the atom-atom quantum correlations along all the three mutually orthogonal directions, that is, along  $x^\prime$, $y^\prime$ and $z^\prime$ directions. It is to be mentioned here that, in Ref. \cite{Ram1} and \cite{Ram2} we took $S$ as the mean squared value of $C_X$ and $C_Y$. In that case $S$ takes the dimension of $\hat{J}^4$. Now, in this paper, we define $S$ in such a way that, it has the dimension of $\hat{J}^2$. For that, we define 
$S$ as the square root of the mean squared value of $C_X$, $C_Y$ and 
$C_Z$. We also incorporate the correlation in the $\hat{J}_{z^\prime}$ 
quadrature in our definition of $S$ for the sake of completeness, that is to construct a total measure of all atom-atom correlations.

Now, using Eqs. (\ref{2.10a1}), (\ref{2.10b1}), and
(\ref{2.9a4}) we can express $S$ as,
\begin{eqnarray}
S &=& \frac{1}{\sqrt{3}}\Big{[}\Delta {J_{x^\prime}}^2\Big{(}\Delta {J_{x^\prime}}^2 -\frac{N}{2}\Big{)} + \Delta {J_{y^\prime}}^2\Big{(}\Delta {J_{y^\prime}}^2-\frac{N}{2}\Big{)}\nonumber\\ 
&+& \Delta J_{z^\prime}^2\bigg{(}\Delta J_{z^\prime}^2 - \frac{N}{2} +\frac{2}{N}|\langle{\bf \hat{J}}\rangle|^2\bigg{)}\nonumber\\
 &+& \bigg{(}
\frac{1}{N}|\langle{\bf \hat{J}}\rangle|^2 - \frac{N}{4}\bigg{)}^2
+ \frac{N^2}{8}\Big{]}^{1/2}.
\label{2.12}
\end{eqnarray}
Now, if $\xi_{R_{x}}$, $\xi_{R_{y}}$ and $\xi_{R_{z}}$ are the spectroscopic squeezing parameters used in Ramsey spectroscopy \cite{Wineland}, given as
\begin{equation}
\xi_{R_{x}} = \frac{\sqrt{2j}}{|\langle{\bf \hat{J}}\rangle|}\Delta J_{x^\prime},~~\xi_{R_{y}} = \frac{\sqrt{2j}}{|\langle{\bf \hat{J}}\rangle|}
\Delta J_{y^\prime},~~ \xi_{R_{z}} = \frac{\sqrt{2j}}{|\langle{\bf \hat{J}}\rangle|}\Delta J_{z^\prime}
\label{2.13}
\end{equation}
then,
\begin{eqnarray}
S &=& \frac{1}{\sqrt{3}}\Bigg[ \frac{\xi_{R_x}^2|\langle\hat{\mathbf J}\rangle|^2}{2j} 
\bigg( \frac{\xi_{R_x}^2|\langle\hat{\mathbf J}\rangle|^2}{2j} - 
\frac{N}{2} \bigg) +  
\frac{\xi_{R_y}^2|\langle\hat{\mathbf J}\rangle|^2}{2j}\nonumber\\ 
&\times& \bigg(\frac{\xi_{R_y}^2|\langle\hat{\mathbf J}\rangle|^2}{2j} - 
\frac{N}{2} \bigg) + \frac{\xi_{R_z}^2|\langle\hat{\mathbf J}\rangle|^2}{2j}\bigg{(}\frac{\xi_{R_z}^2|\langle\hat{\mathbf J}\rangle|^2}{2j} - \frac{N}{2}\nonumber\\
&+& \frac{2}{N}|\langle\hat{\mathbf J}\rangle|^2 \bigg{)} +
\bigg{(}\frac{1}{N}|\langle{\bf \hat{J}}\rangle|^2 - \frac{N}{4}\bigg{)}^2
+ \frac{N^2}{8}\Bigg]^{1/2}.
\label{2.14}
\end{eqnarray}
Thus, the quantum correlation parameter $S$ is connected 
to the experimentally measurable quantities. 

Now, if all the atoms in the assembly are uncorrelated, that is, the quantum state $|\Psi\rangle$ is a separable state, then all the conditions in Eqs. (\ref{2.5d24}) are satisfied, and hence, $C_X$, $C_Y$, and $C_Z$
become zero, producing $S = 0$. If the atoms in the assembly are correlated, then, all the conditions in Eqs. (\ref{2.5d24}) are not satisfied, and as a consequence, either $C_X$ or, $C_Y$ or, $C_Z$, or any two of the three, or, all are non-zero. In that case $S \ne 0$. Therefore, the non-zero value of $S$ implies the presence of quantum correlations among the atoms.
Now, we can take the numerical value of 
$S$ as a total measure of the amount of quantum correlation present among the $N$ two-level atoms. Thus, by measuring $S$ with the help of 
Eqs. (\ref{2.12}) or (\ref{2.14}) for a quantum state of $N$ two-level atoms we can measure the net amount of quantum correlation present in that system. In the following section we apply this idea to measure the net amount of quantum correlation present in an atomic Schr\"{o}dinger cat state, which is of considerable interest in the conceptual foundations of quantum mechanics.

\section{III. Quantification of quantum correlations in atomic 
Schr\"{O}dinger cat states}

We consider a system of $N$ identical two-level atoms each of transition frequency $\omega_0$ interacting collectively with a single mode electromagnetic field in a cavity whose characteristic frequency nearest to 
$\omega_0$ is $\omega_c$. The cavity temperature is represented by the average number of thermal photons $\bar{n}$ present in the cavity. The rate of loss of photons is given as $2\kappa$, where $\kappa$ is the cavity bandwidth. We consider the cavity to be highly detuned, that is, $\delta_c = \omega_c - \omega_0$ is very large such that the condition 
$|i\delta_c + \kappa| >> g\sqrt{N}$ (where $g$ is the atom-photon coupling constant) is satisfied. We assume that the quality factor $Q$ of the cavity is very large making $\kappa$ very small, but we treat the cavity as a dispersive type. The effective Hamiltonian describing the dynamics of the atoms in the cavity takes the form \cite{Agarwal1}
\begin{equation}
H_{eff} = \hbar\eta \Big{[}{\hat{J}}^2 - {\hat{J}_z}^2 + (2\bar{n} + 1)\hat{J}_z\Big{]},
\label{3.1}
\end{equation}
where 
\begin{equation}
\eta = \frac{g^2\delta_c}{\kappa^2 + {\delta_c}^2}.
\label{3.2}
\end{equation}
It can be noted that the temperature dependent term corresponds to a simple rotation and therefore, we drop it in further considerations, that is, we set $\bar{n} = 0$. Therefore, the effective Hamiltonian reduces to
\begin{equation}
H = \hbar\eta \Big{[}{\hat{J}}^2 - {\hat{J}_z}^2 + \hat{J}_z\Big{]}.
\label{3.3}
\end{equation}
Now, the unitary time evolution generated by the above Hamiltonian transforms an atomic coherent state into atomic Schr\"{o}dinger cat states at special times \cite{Agarwal1}.

Let initially a system of $N$ two-level atoms are prepared in an atomic coherent state \cite{Arecchi},
\begin{eqnarray}
|\psi(0)\rangle &\equiv& |\theta,\phi\rangle = \sum_{k=0}^{N}\sqrt{\frac{N!}{(N-k)!k!}}\exp{(ik\phi)}\nonumber\\
&\times& \sin^{N-k}\Big{(}\frac{\theta}{2}\Big{)}\cos^k\Big{(}\frac{\theta}{2}
\Big{)} \Big{|}\frac{N}{2} - k\Big{\rangle}.
\label{3.4}
\end{eqnarray}
If the above state is given a time evolution by the Hamiltonian $H$ in Eq. (\ref{3.3}), we obtain
\begin{eqnarray}
&&|\psi(t)\rangle = e^{-iHt/\hbar}|\theta,\phi\rangle = \sum_{k=0}^{N}\sqrt{\frac{N!}{(N-k)!k!}}\exp{(ik\phi)}\nonumber\\
&\times&\sin^{N-k}\Big{(}\frac{\theta}{2}\Big{)}\cos^k\Big{(}\frac{\theta}{2}
\Big{)}\nonumber\\
&\times&\exp{[-i\tau\{ N + (N-1)k - k^2\}]}~\Big{|}\frac{N}{2} - 
k\Big{\rangle}, 
\label{3.5}
\end{eqnarray} 
where $\tau = \eta t$.
Now, it has been shown in Ref. \cite{Agarwal1}, that at special times 
$\tau = \frac{\pi}{m}$, where $m$ is an integer, the above state can be expressed as superposition of several atomic coherent states as
\begin{eqnarray}
|\psi(t)\rangle &=& e^{-iHt/\hbar}|\theta,\phi\rangle = \exp\Big{[}-i\frac{\pi N}{m}\Big{]}\nonumber\\
&\times&\sum_{q=0}^{m-1} f_{q}^{(o)}\Big{|}\theta, \phi+\pi\frac{2q-N}{m}\Big{\rangle}
\label{3.6}
\end{eqnarray}
for $m$ odd, and
\begin{eqnarray}
|\psi(t)\rangle &=& e^{-iHt/\hbar}|\theta,\phi\rangle = \exp\Big{[}-i\frac{\pi N}{m}\Big{]}\nonumber\\
&\times&\sum_{q=0}^{m-1} f_{q}^{(e)}\Big{|}\theta, \phi+\pi\frac{2q-N+1}{m}\Big{\rangle}
\label{3.7}
\end{eqnarray}
for $m$ even. Here, $f_{q}^{(o)}$ and $f_(q)^{(e)}$ are given as \cite{Agarwal1},
\begin{eqnarray}
f_{q}^{(o)} = \frac{1}{m}\sum_{k=0}^{m-1}\exp\Big{[}-i\frac{2\pi q}{m}k\Big{]}
\exp\Big{[}i\frac{\pi}{m}k(k+1)\Big{]}
\label{3.8}
\end{eqnarray}
and
\begin{eqnarray}
f_{q}^{(e)} = \frac{1}{m}\sum_{k=0}^{m-1}\exp\Big{[}-i\frac{2\pi q}{m}k\Big{]}
\exp\Big{[}i\frac{\pi}{m}k^2\Big{]}.
\label{3.9}
\end{eqnarray}
We observe from Eqs. (\ref{3.6}) and (\ref{3.7}), that the various atomic coherent states in the superposition differ by phase and the state 
$|\psi(t)\rangle$ is an atomic Schr\"{o}dinger cat state. It has been shown in Ref. \cite{Agarwal1}, that the state $|\psi(t)\rangle$ shows interesting interference patterns in its quasidistributions. Now, 
$|\psi(t)\rangle$ cannot be expressed as a product states of individual spins and hence atom-atom correlations are important in this cat state.  
This is our motivation to quantify the amount of atomic correlations present in this state.

We now proceed to calculate the amount of quantum correlations present in the above state. To obtain this we calculate the average values of the collective atomic operators $\hat{J}_x$, $\hat{J}_y$ and $\hat{J}_z$, for the state $|\psi(t)\rangle$ (with $N =2j$).
Now, it can be seen that for the initial state $|\psi(0)\rangle = |\theta,\phi\rangle$, we have
\begin{eqnarray}
\langle\theta,\phi|\hat{J}_x|\theta,\phi\rangle &=& j\sin\theta\cos\phi,\label{3.9a1}\\
\langle\theta,\phi|\hat{J}_y|\theta,\phi\rangle &=& j\sin\theta\sin\phi,\label{3.9a2}\\
\langle\theta,\phi|\hat{J}_z|\theta,\phi\rangle &=& -j\cos\theta.
\label{3.9a3}
\end{eqnarray}
Now, for the state 
\begin{eqnarray}
|\psi(t)\rangle &=& \exp(-iH_{eff}t/\hbar)|\theta,\phi\rangle\nonumber\\
&=& \exp{-i\tau \Big{[}{\hat{J}}^2 - {\hat{J}_z}^2 + (2\bar{n} + 1)\hat{J}_z\Big{]}}|\theta,\phi\rangle,
\label{3.9a4} 
\end{eqnarray}
we calculate the average values $\langle\hat{J}_x\rangle$, 
$\langle\hat{J}_y\rangle$, and $\langle\hat{J}_z\rangle$, which are as shown below.
\begin{eqnarray}
\langle\psi(t)|\hat{J}_x|\psi(t)\rangle &=& j \sin\theta (T_1)^{j-1}\Big{[}
\cos^2\Big{(}\frac{\theta}{2}\Big{)}\cos(2j\Theta_1 - \phi)\nonumber\\
 &+& \sin^2\Big{(}\frac{\theta}{2}\Big{)}\cos(2j\Theta_1
- 2\tau -\phi)\Big{]},
\label{3.10}
\end{eqnarray}
\begin{eqnarray}
\langle\psi(t)|\hat{J}_y|\psi(t)\rangle &=& -j \sin\theta (T_1)^{j-1}\Big{[}
\cos^2\Big{(}\frac{\theta}{2}\Big{)}\sin(2j\Theta_1 - \phi)\nonumber\\
 &+& \sin^2\Big{(}\frac{\theta}{2}\Big{)}\sin(2j\Theta_1
- 2\tau -\phi)\Big{]}
\label{3.11}
\end{eqnarray}
where,
\begin{equation}
T_1 = \sin^4\Big{(}\frac{\theta}{2}\Big{)} + \frac{1}{2}\sin^2\theta\cos 2\tau +
\cos^4\Big{(}\frac{\theta}{2}\Big{)}
\label{3.12}
\end{equation}
and
\begin{equation}
\Theta_1 = \tan^{-1}\Big{[}-\tan\tau\cos\theta\Big{]}.
\label{3.13}
\end{equation}
Now,
\begin{equation}
\langle\psi(t)|\hat{J}_z|\psi(t)\rangle = -j \cos\theta.
\label{3.14}
\end{equation}
We note from the above results that, the mean spin vector 
$\langle{\bf\hat{J}}\rangle$ 
points in an arbitrary direction in space. We, therefore, rotate the coordinate system $\{x, y, z\}$ to $\{x^\prime, y^\prime, z^\prime\}$, such that the mean spin vector points along the $z^\prime$-axis and hence, the components of ${\bf\hat{J}}$ in the rotated frame become
\begin{eqnarray}
\hat{J}_{x^\prime}&=&\hat{J}_x\cos\phi_1\cos\theta_1+\hat{J}_y\sin\phi_1\cos\theta_1\nonumber\\
&-& \hat{J}_z\sin\theta_1\label{3.15}\\
\hat{J}_{y^\prime}&=&-\hat{J}_x\sin\phi_1+\hat{J}_y\cos\phi_1\label{3.16}\\
\hat{J}_{z^\prime}&=&\hat{J}_x\cos\phi_1\sin\theta_1+\hat{J}_y\sin\phi_1\sin\theta_1\nonumber\\
&+& \hat{J}_z\cos\theta_1\label{3.17}
\end{eqnarray}
where,
\begin{eqnarray}
\cos\theta_1 &=& \frac{\langle\hat{J}_z\rangle}{|\langle{\bf\hat{J}}\rangle|}, ~~\sin\theta_1 = \frac{\sqrt{\langle\hat{J}_x\rangle^2+\langle\hat{J}_y\rangle^2}}{|\langle{\bf\hat{J}}\rangle|},\label{3.18}\\
\cos\phi_1 &=& \frac{\langle\hat{J}_x\rangle}{\sqrt{\langle\hat{J}_x\rangle^2+\langle\hat{J}_y\rangle^2}},\label{3.19}\\
 \sin\phi_1 &=& \frac{\langle\hat{J}_y\rangle}{\sqrt{\langle\hat{J}_x\rangle^2+\langle\hat{J}_y\rangle^2}}.
\label{3.20}
\end{eqnarray}
We can check that 
\begin{equation}
\langle\hat{J}_{x^\prime}\rangle = \langle\hat{J}_{y^\prime}\rangle = 0
\end{equation}
and hence, the mean spin vector lies along the $z^\prime$-axis.
We, now, calculate
\begin{equation}
\Delta J_{x^\prime,y^\prime,z^\prime}^2 = \langle\hat{J}_{x^\prime,y^\prime,z^\prime}^2\rangle - \langle\hat{J}_{x^\prime,y^\prime,z^\prime}\rangle^2.
\label{3.21}
\end{equation}
For that we need $\langle\hat{J}_x^2\rangle$, $\langle\hat{J}_y^2\rangle$,
$\langle\hat{J}_z^2\rangle$, $\langle\hat{J}_x\hat{J}_y + \hat{J}_y
\hat{J}_x\rangle$, $\langle\hat{J}_y\hat{J}_z + \hat{J}_z
\hat{J}_y\rangle$ and $\langle\hat{J}_x\hat{J}_z + \hat{J}_z
\hat{J}_x\rangle$.
We present the final results of the above values as
\begin{eqnarray}
\langle\hat{J}_x^2\rangle &=& \frac{j(2j-1)}{4}\sin^2\theta (T_2)^{j-1}
\cos\Big{[}(6-4j)\tau + 2\phi\nonumber\\
&+& (2j-2)\Theta_2\Big{]}
+ \frac{j(j+1)}{2} - \frac{1}{2}\Big{[}j^2\sin^4\Big{(}\frac{\theta}{2}{)}
\nonumber\\
&-&\frac{1}{2}j^2\sin^2\theta + \frac{1}{2}j\sin^2\theta + j^2\cos^4\Big{(}\frac{\theta}{2}\Big{)}\Big{]},
\label{3.22}
\end{eqnarray}
\begin{eqnarray}
\langle\hat{J}_y^2\rangle &=& -\frac{j(2j-1)}{4}\sin^2\theta (T_2)^{j-1}
\cos\Big{[}(6-4j)\tau + 2\phi\nonumber\\
&+& (2j-2)\Theta_2\Big{]}
+ \frac{j(j+1)}{2} - \frac{1}{2}\Big{[}j^2\sin^4\Big{(}\frac{\theta}{2}{)}
\nonumber\\
&-&\frac{1}{2}j^2\sin^2\theta + \frac{1}{2}j\sin^2\theta + j^2\cos^4\Big{(}\frac{\theta}{2}\Big{)}\Big{]},
\label{3.23}
\end{eqnarray}
\begin{eqnarray}
&&\langle\hat{J}_x\hat{J}_y + \hat{J}_y\hat{J}_x\rangle = \frac{j(2j-1)}{2}
\sin^2\theta (T_2)^{j-1}\nonumber\\
&\times&\sin\Big{[}(6-4j)\tau + 2\phi + (2j-2)\Theta_2\Big{]},
\label{3.24}
\end{eqnarray}
where
\begin{equation}
T_2 = \sin^4\Big{(}\frac{\theta}{2}\Big{)} + \frac{1}{2}\sin^2\theta\cos 4\tau + \cos^4\Big{(}\frac{\theta}{2}\Big{)}
\label{3.25}
\end{equation}
and
\begin{equation}
\Theta_2 = \tan^{-1}\bigg{[}\frac{\cos^2(\theta/2)\sin 4\tau}{\sin^2(\theta/2)+\cos^2(\theta/2)\cos 4\tau}\bigg{]}.
\label{3.26}
\end{equation}
Now,
\begin{equation}
\langle\hat{J}_z^2\rangle = \frac{1}{2}j\sin^2\theta + j^2\Big{[}\sin^4\Big{(}\frac{\theta}{2}\Big{)} -\frac{1}{2}\sin^2\theta + \cos^4\Big{(}\frac{\theta}{2}\Big{)} \Big{]},
\label{3.27}
\end{equation}
\begin{eqnarray}
&&\langle\hat{J}_y\hat{J}_z + \hat{J}_z\hat{J}_y\rangle = j(2j-1)
\sin\theta (T_1)^{j-1}\bigg{[}\Big{\{}\sin^2\Big{(}\frac{\theta}{2}\Big{)}\nonumber\\ 
&-& \cos^2\Big{(}\frac{\theta}{2}\Big{)}\cos 2\tau\Big{\}}\sin\Big{\{}(1-j)2\tau + \phi + (2j-2)\Theta_3\Big{\}}\nonumber\\ 
&-& \cos^2\Big{(}\frac{\theta}{2}\Big{)}\sin 2\tau\cos\Big{\{}(1-j)2\tau 
+ \phi + (2j-2)\Theta_3\Big{\}}\bigg{]}\nonumber\\
\label{3.28}
\end{eqnarray}
and
\begin{eqnarray}
&&\langle\hat{J}_x\hat{J}_z + \hat{J}_z\hat{J}_x\rangle = j(2j-1)
\sin\theta (T_1)^{j-1}\bigg{[}\Big{\{}\sin^2\Big{(}\frac{\theta}{2}\Big{)}\nonumber\\ 
&-& \cos^2\Big{(}\frac{\theta}{2}\Big{)}\cos 2\tau\Big{\}}\cos\Big{\{}(1-j)2\tau + \phi + (2j-2)\Theta_3\Big{\}}\nonumber\\ 
&+& \cos^2\Big{(}\frac{\theta}{2}\Big{)}\sin 2\tau\sin\Big{\{}(1-j)2\tau 
+ \phi + (2j-2)\Theta_3\Big{\}}\bigg{]},\nonumber\\
\label{3.29}
\end{eqnarray}
where
\begin{equation}
\Theta_3 = \tan^{-1}\bigg{[}\frac{\cos^2(\theta/2)\sin 2\tau}{\sin^2(\theta/2)+\cos^2(\theta/2)\cos 2\tau}\bigg{]}.
\label{3.30}
\end{equation}
Now, using the expressions of $\cos\theta_1$, $\sin\theta_1$, $\cos\phi_1$ and $\sin\phi_1$, given in Eqs. (\ref{3.18}), (\ref{3.19}) and (\ref{3.20}) in Eqs. (\ref{3.15}), (\ref{3.16}) and (\ref{3.17}) and using Eq. (\ref{3.21}), we obtain the expressions of $\Delta J_{x^\prime}^2$, 
$\Delta J_{y^\prime}^2$ and $\Delta J_{z^\prime}^2$ respectively as
\begin{eqnarray}
\Delta J_{x^\prime}^2 &=& \frac{1}{|\langle{\bf\hat{J}}\rangle|^2\Big{(}\langle\hat{J}_x\rangle^2
+ \langle\hat{J}_y\rangle^2\Big{)}}\Bigg{[}\langle\hat{J}_x^2\rangle
\langle\hat{J}_x\rangle^2\langle\hat{J}_z\rangle^2 + \langle\hat{J}_y^2\rangle
\nonumber\\
&\times&\langle\hat{J}_y\rangle^2
\langle\hat{J}_z\rangle^2 + \langle\hat{J}_z^2\rangle \Big{(}\langle\hat{J}_x\rangle^2 + \langle\hat{J}_y\rangle^2\Big{)}^2 \nonumber\\ 
&+& 
\langle\hat{J}_x\hat{J}_y + \hat{J}_y\hat{J}_x\rangle
\langle\hat{J}_x\rangle\langle\hat{J}_y\rangle\langle\hat{J}_z\rangle^2 - \Big{(}\langle\hat{J}_x\rangle^2 + \langle\hat{J}_y\rangle^2\Big{)}\nonumber\\
&\times&\Big{\{}
\langle\hat{J}_x\hat{J}_z + \hat{J}_z\hat{J}_x\rangle
\langle\hat{J}_x\rangle\langle\hat{J}_z\rangle\nonumber\\
&+& \langle\hat{J}_y\hat{J}_z + \hat{J}_z\hat{J}_y\rangle\langle\hat{J}_y\rangle\langle\hat{J}_z\rangle\Big{\}} \Bigg{]},
\label{3.31}
\end{eqnarray}
\begin{eqnarray}
\Delta J_{y^\prime}^2 &=& \frac{1}{\Big{(}\langle\hat{J}_x\rangle^2
+ \langle\hat{J}_y\rangle^2\Big{)}}\Bigg{[}\langle\hat{J}_x^2\rangle\langle\hat{J}_y\rangle^2 + \langle\hat{J}_y^2\rangle\langle\hat{J}_x\rangle^2\nonumber\\
 &-& \langle\hat{J}_x\hat{J}_y + \hat{J}_y\hat{J}_x\rangle\langle\hat{J}_x\rangle \langle\hat{J}_y\rangle\Bigg{]}
\label{3.32}
\end{eqnarray}
and
\begin{eqnarray}
\Delta J_{z^\prime}^2 &=& \frac{1}{|\langle{\bf\hat{J}}\rangle|^2}\Bigg{[}\Delta J_x^2\langle\hat{J}_x\rangle^2 + \Delta J_y^2\langle\hat{J}_y\rangle^2 + \Delta J_z^2\langle\hat{J}_z\rangle^2\nonumber\\
&+& \Big{(}\langle\hat{J}_x\hat{J}_y + \hat{J}_y\hat{J}_x\rangle - 2\langle\hat{J}_x\rangle\langle\hat{J}_y\rangle\Big{)}\langle\hat{J}_x\rangle\langle\hat{J}_y\rangle\nonumber\\
&+& \Big{(}\langle\hat{J}_y\hat{J}_z + \hat{J}_z\hat{J}_y\rangle - 2\langle\hat{J}_y\rangle\langle\hat{J}_z\rangle\Big{)}\langle\hat{J}_y\rangle\langle\hat{J}_z\rangle\nonumber\\
&+& \Big{(}\langle\hat{J}_x\hat{J}_z + \hat{J}_z\hat{J}_x\rangle - 2\langle\hat{J}_x\rangle\langle\hat{J}_z\rangle\Big{)}\langle\hat{J}_x\rangle\langle\hat{J}_z\rangle\Bigg{]}.
\label{3.33}
\end{eqnarray}
We can now quantify the amount of quantum correlation in the cat state by using Eqs. (\ref{2.11}) or (\ref{2.12}) and Eqs. (\ref{3.31}), (\ref{3.32}) and (\ref{3.33}) by using the various types of average values of the atomic operators and their combinations as calculated above. We observe from Eqs. (\ref{3.6}) and (\ref{3.7}), that for various values of $m$, that is for various interaction times $\tau =\pi/m$, we have various superpositions of atomic coherent states and thus, we obtain various atomic cat states. In Table I we have shown the quantum correlations along the three mutually orthogonal directions, that is, $x^\prime$, $y^\prime$ and $z^\prime$ directions of the system of atoms and also the net quantum correlation among the atoms, represented by the magnitude of $S$ for various atomic cat states corresponding to various values of $m$, that is, for various interaction times $\tau$. These results are for, total number of atoms $N = 10$, $\theta=\pi/4$, that is when the number of atoms in the excited state is one-fourth of the total number of atoms in the assembly, and $\phi =0$. 

\begin{table}[ht]
\caption{Values of $C_X$, $C_Y$, $C_Z$ and $S$ for various values of 
$m$, for the number of atoms $N = 10$}
\centering
\begin{tabular}{c c c c c}
\hline\hline
m & $C_X$ & $C_Y$ & $C_Z$ & $S$ \\ [0.5ex]
 2 & -0.06579 &  11.25000 &      0.04382 &  6.49535\\    
 3 &  4.76608  &     6.05441 &     0.26581 &      4.45129\\    
 4 &  3.94977  &     5.59583 &     0.85970 &      3.98552\\    
 5 &  2.67068  &     5.20093 &      1.33710 &      3.46266\\    
 6 &  1.63697  &     4.77206 &      1.45857 &      3.03203\\    
 7 & 0.95262  &     4.32450 &      1.35167 &      2.67306\\    
 8 & 0.52939  &     3.88241 &      1.16369 &      2.35991\\    
 9 & 0.27321  &     3.46535 &     0.97139 &      2.08382\\    
 10 &  0.11898 &      3.08491 &     0.80294 &      1.84196\\
 [1 ex]   
\hline
\end{tabular}
\label{table:quant}
\end{table}

Now, in Fig. 1, we show the variation of the amount of quantum correlation, that is, the value of $S$ with the number of atoms $N$ varying from $N = 2$ to 
$N = 10$. We show three curves corresponding to three values of the interaction time $\tau$, which are $\tau = \pi/8, \pi/6$ and $\pi/4$. We observe interesting feature for $\tau = \pi/6$. We note from Fig. 1, that the amount of quantum correlation for odd number of atoms is greater than its next higher even number of atoms. That is the quantum correlation ($S$) for $N =3$ is higher than that for $N = 4$. Similarly, $S$ for $N=5$ is higher than $S$ for $N = 6$. $S$ for $N = 7$ is higher than $S$ for $N = 8$ and $S$ for $N = 9$ is higher than $S$ for $N = 10$. This interesting feature decreases as we deviate from the interaction time $\tau = \pi/6$. For the interaction time $\tau = \pi/8$ the amount of quantum correlation increases smoothly with the increase in the number of atoms $N$, but for $\tau = \pi/4$, it increases in a zig-zag fashion with the increase in the number of atoms.

In Fig. 2, we show the variation of $S$ with $N$, where $N$ varies from $N = 10$ to $N = 20$. We note that, the increase in $S$ with $N$ is smooth for 
$\tau = \pi/8$ and $\pi/4$, whereas for $\tau = \pi/6$, the increase in $S$ with $N$ shows slight zig-zag fashion in the begining and becomes almost smooth after 
$N =16$.

In Fig. 3, we show the same curves for $N$ varying from $N = 20$ to $N = 100$.
We note that the variation in $S$ with $N$, for all three interaction times 
($\tau = \pi/8, \pi/6$ and $\pi/4$) become almost same after $N =80$.

\begin{figure}
\begin{center}
\includegraphics[width=6cm, angle=-90]{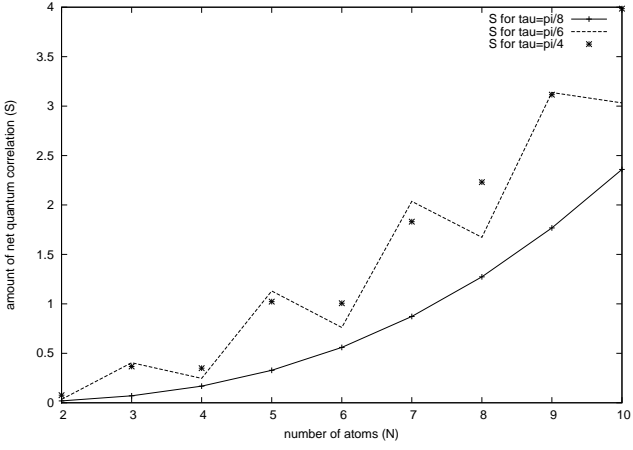}
\caption{Variation of the amount of net quantum correlation ($S$) with the number of atoms ($N$), where $N$ varies from $N=2$ to $N=10$. The three curves are for three values of $\tau$, which are, $\tau=\pi/8$, $\pi/6$ and $\pi/4$. We have taken $\theta = \pi/4$ and $\phi = 0$ in all the three cases.}
\end{center}
\label {fig1}
\end{figure}

\begin{figure}
\begin{center}
\includegraphics[width=6cm, angle = -90]{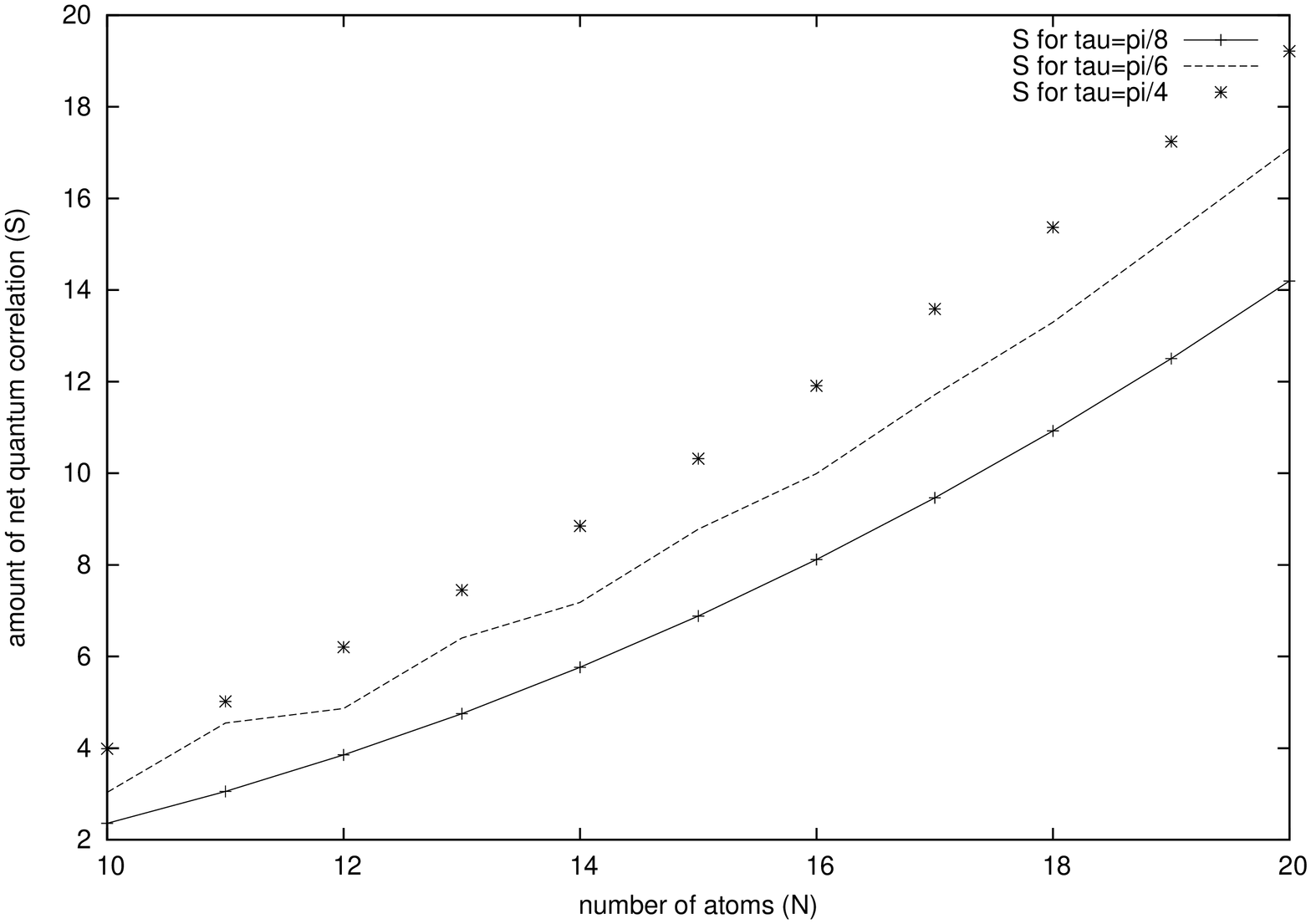}
\caption{Variation of the amount of net quantum correlation ($S$) with the number of atoms ($N$), where $N$ varies from $N=10$ to $N=20$. The three curves are for three values of $\tau$, which are, $\tau=\pi/8$, $\pi/6$ and $\pi/4$. We have taken $\theta = \pi/4$ and $\phi = 0$ in all the three cases.}
\end{center}
\label {fig2}
\end{figure}

\begin{figure}
\begin{center}
\includegraphics[width=6cm, angle = -90]{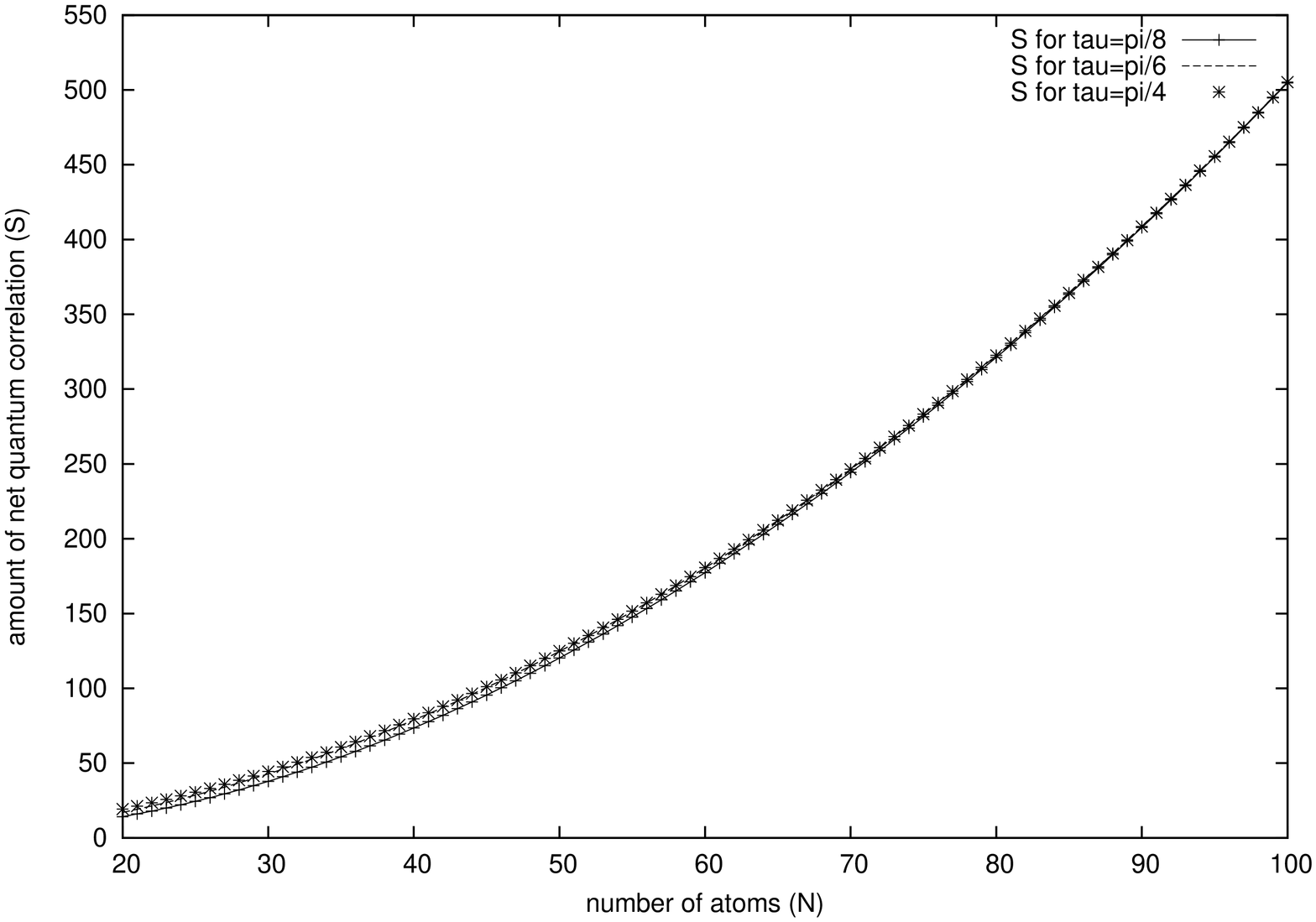}
\caption{Variation of the amount of net quantum correlation ($S$) with the number of atoms ($N$), where $N$ varies from $N=20$ to $N=100$. The three curves are for three values of $\tau$, which are, $\tau=\pi/8$, $\pi/6$ and $\pi/4$. We have taken $\theta = \pi/4$ and $\phi = 0$ in all the three cases.}
\end{center}
\label {fig3}
\end{figure}

\begin{figure}
\begin{center}
\includegraphics[width=6cm, angle = -90]{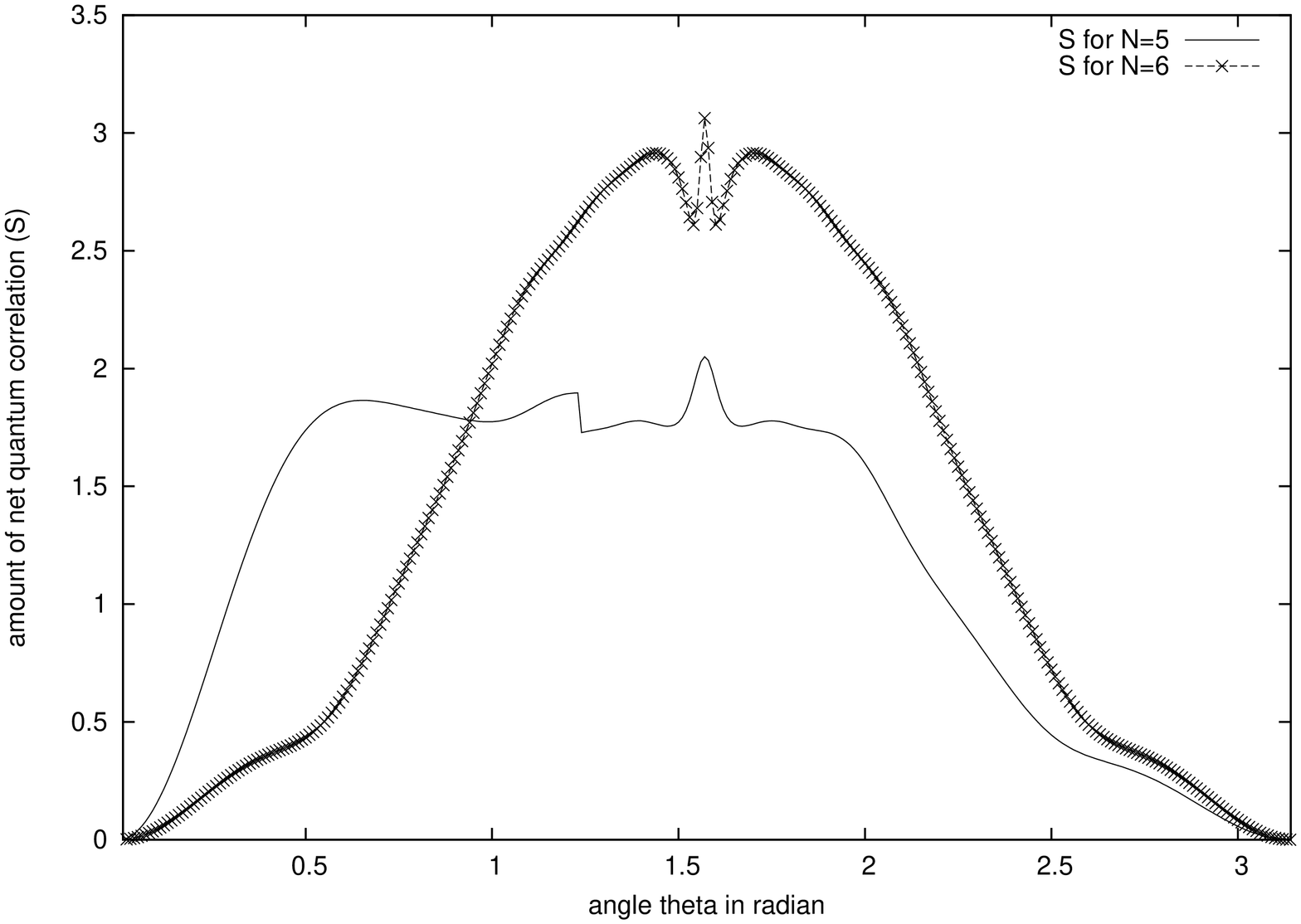}
\caption{Variation of the amount of net quantum correlation ($S$) with the angle 
$\theta$, in radian, where $\theta$ varies from $\theta =(\pi/314)$ to $\theta = \pi$ radian. The two curves are for two values of $N$, which are, 
 $N=5$ and $N=6$. We have taken $\tau=\pi/3$ and $\phi = 0$ in both the two cases.}
\end{center}
\label {fig4}
\end{figure}

\begin{figure}
\begin{center}
\includegraphics[width=6cm, angle = -90]{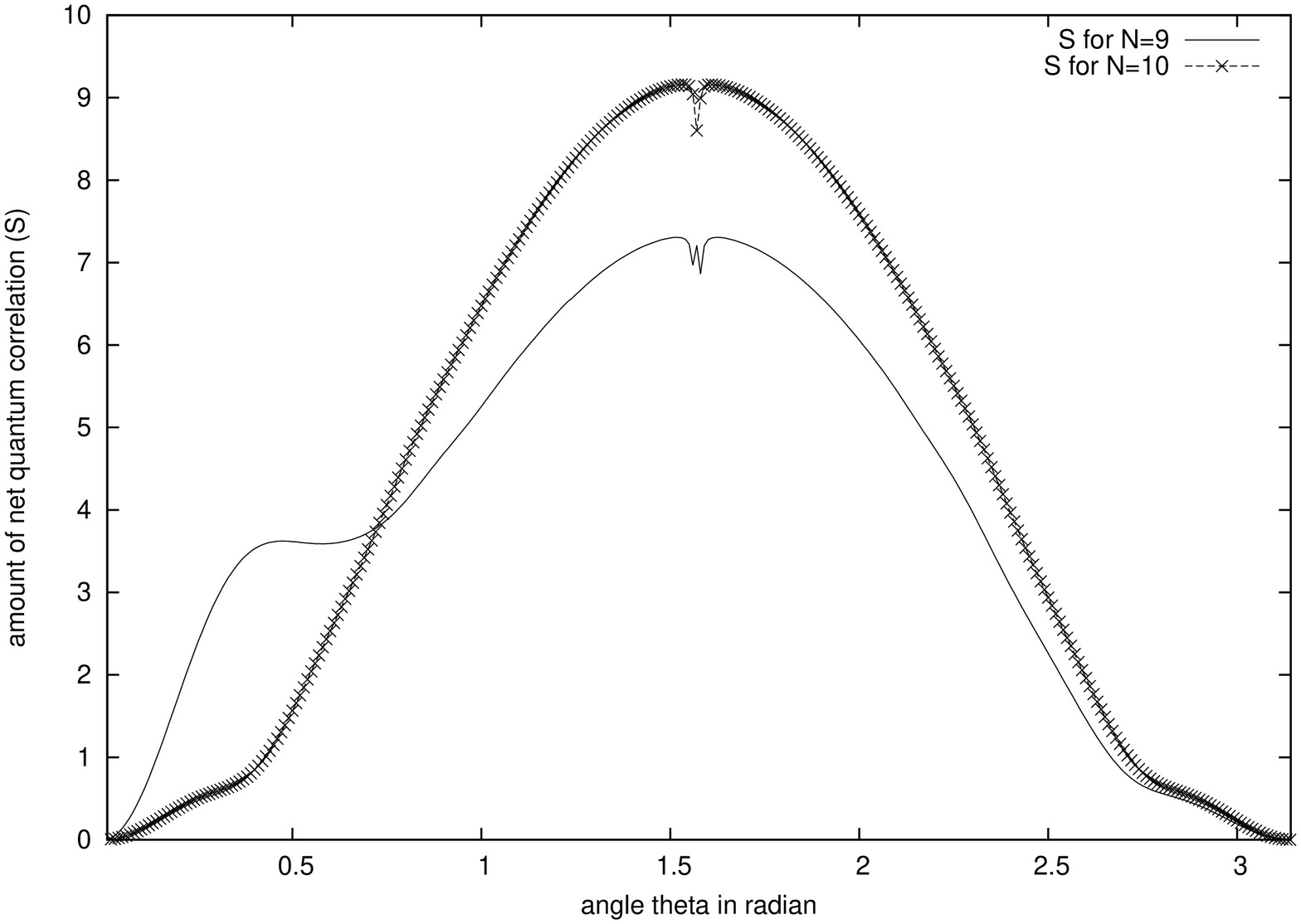}
\caption{Variation of the amount of net quantum correlation ($S$) with the angle 
$\theta$, in radian, where $\theta$ varies from $\theta = (\pi/314)$ to $\theta = \pi$ radian. The two curves are for two values of $N$, which are, 
 $N=9$ and $N=10$. We have taken $\tau = \pi/3$ and $\phi = 0$ in both the two cases.}
\end{center}
\label {fig5}
\end{figure}

\begin{figure}
\begin{center}
\includegraphics[width=6cm, angle = -90]{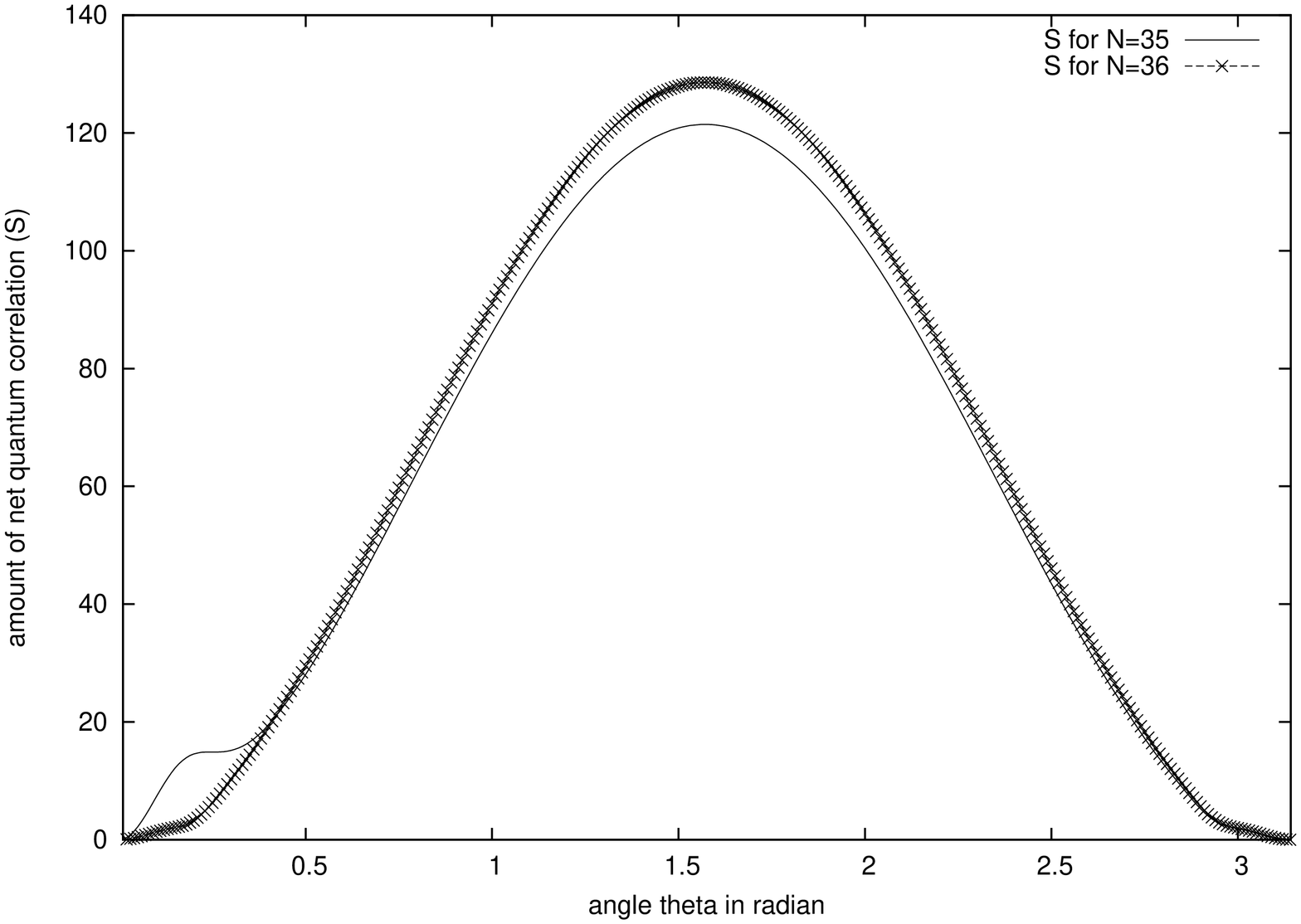}
\caption{Variation of the amount of net quantum correlation ($S$) with the angle $\theta$, in radian, where $\theta$ varies from $\theta = (\pi/314)$ to $\theta = \pi$ radian. The two curves are for two values of $N$, which are, 
 $N=35$ and $N=36$. We have taken $\tau = \pi/3$ and $\phi = 0$ in both the two cases.}
\end{center}
\label {fig6}
\end{figure}

\begin{figure}
\begin{center}
\includegraphics[width=6cm, angle = -90]{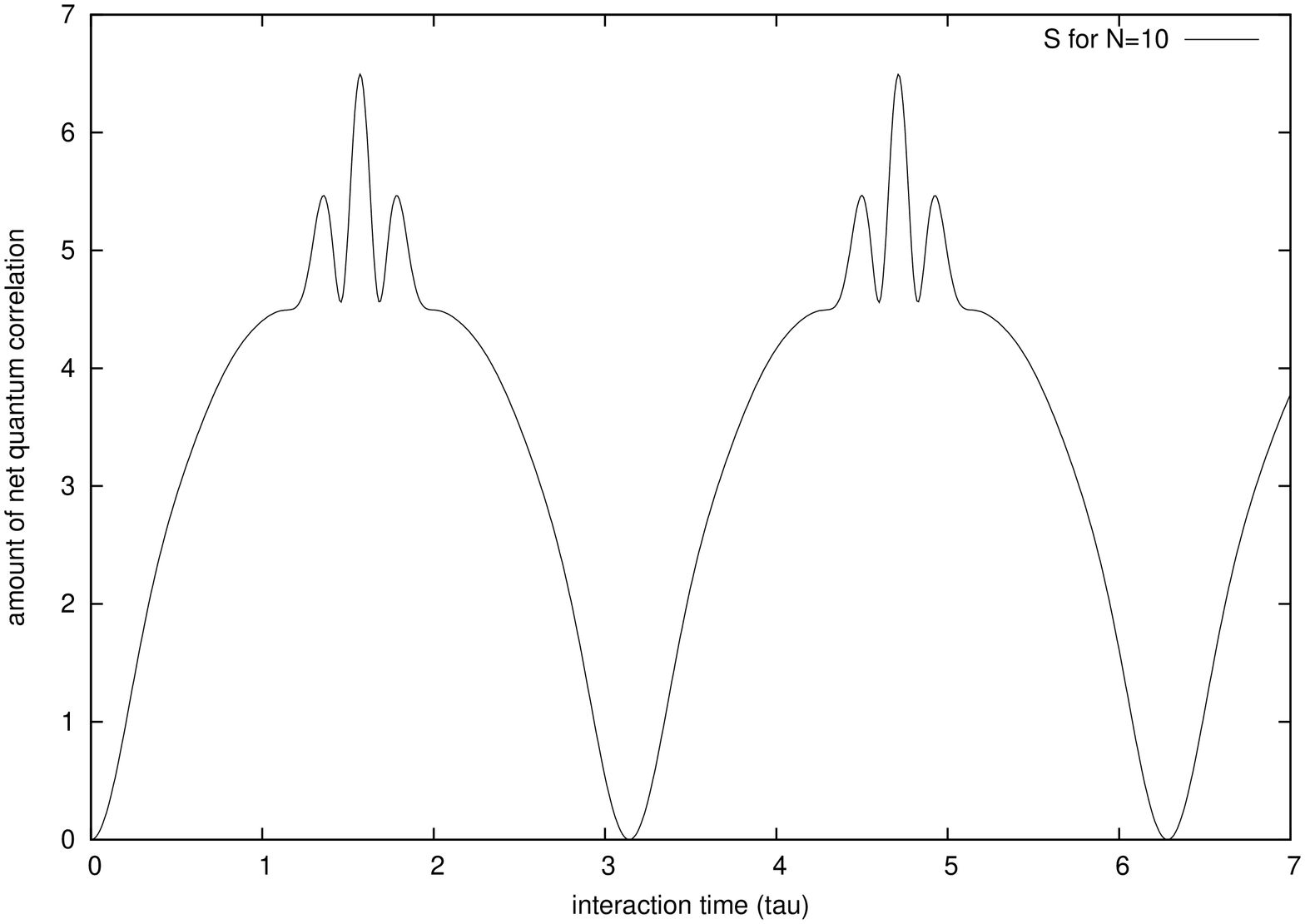}
\caption{Variation of the amount of net quantum correlation ($S$) with the interaction time $\tau$. We have taken $N = 10$, $\theta = \pi/4$ and $\phi = 0$. }
\end{center}
\label {fig7}
\end{figure}

In Fig. 4, 5 and 6 we show the variation of $S$ with the angle $\theta$ in radian, for different values of $N$.
 The value of $\theta$ gives a measure of the number of atoms in the excited states in the initial atomic coherent state $|\theta,\phi\rangle$.
When $\theta = 0$, we observe from Eq. (\ref{3.4}), that, only the term for $k = N$ in the summation exists and the corresponding state is given as
\begin{equation}
|\psi(0)\rangle = |\theta=0,\phi\rangle = \exp(iN\phi)\Bigg{|}\frac{-N}{2}\Bigg{\rangle},
\label{3.33a1}
\end{equation}
that is, all the spins or equivalently all the two-level atoms are in their lower states.  Using
 Eqs. (\ref{3.9a1}), (\ref{3.9a2}) and 
(\ref{3.9a3}), we have
\begin{eqnarray}
\langle\theta=0,\phi|\hat{J}_x|\theta=0,\phi\rangle &=& 0,\\
\langle\theta=0,\phi|\hat{J}_y|\theta=0,\phi\rangle &=& 0,\\
\langle\theta=0,\phi|\hat{J}_z|\theta=0,\phi\rangle &=& -j = -\frac{N}{2}
\label{3.34}
\end{eqnarray}
 Similarly, when $\theta = \pi$, we observe from Eq. (\ref{3.4}), that, only the term for $k = 0$ in the summation exists and the the corresponding state is given as
\begin{equation}
|\psi(0)\rangle = |\theta=\pi,\phi\rangle = \Bigg{|}\frac{N}{2}\Bigg{\rangle},
\label{3.34a1}
\end{equation} 
that is, all the atoms are in their upper states.
Using Eqs. (\ref{3.9a1}), (\ref{3.9a2}) and (\ref{3.9a3}), we see that
\begin{eqnarray}
\langle\theta=\pi,\phi|\hat{J}_x|\theta=\pi,\phi\rangle &=& 0,\\
\langle\theta=\pi,\phi|\hat{J}_y|\theta=\pi,\phi\rangle &=& 0,\\
\langle\theta=\pi,\phi|\hat{J}_z|\theta=\pi,\phi\rangle &=& j = \frac{N}{2}
\label{3.35}
\end{eqnarray}
 Similarly, $|\theta=\pi/2,\phi\rangle$
is the coherent state of equal number of atoms in their lower states and excited states, when we have
\begin{equation}
\langle\theta=\pi/2,\phi|\hat{J}_z|\theta=\pi/2,\phi\rangle = 0.
\label{3.35a1}
\end{equation}
Therefore, In Fig. 4, we show the variation of the net quantum correlation $S$
 as we increase the number of atoms in the excited states. The two curves are for two values of $N$ which are $N =5$ and $6$. The interaction time has been set equal to $\tau = \pi/3$ and we have taken $\phi = 0$ for simplicity. We observe that when the number of atoms in the assembly is an odd number, the corresponding curve is asymmetric about 
$\theta =\pi/2$ whereas it is symmetric about $\theta = \pi/2$, when the number of atoms in the assembly is an even number. We also abserve that for $N = 5$, the variation of the amount of net quantum correlation ($S$) with $\theta$ shows  irrregular pattern, though it shows maximum at $\theta = \pi/2$, that is, when there are equal number of atoms in the upper and lower levels. 
But for $N =6$, the amount of quantum correlation ($S$) increases gradually with the increase in $\theta$, then it sharply falls before $\theta =\pi/2$ and then sharply increases and attains maximum at $\theta = \pi/2$. Then again it sharply falls just after $\theta = \pi/2$, then sharply increases within a small range of $\theta$ and then gradually decreases to zero at 
$\theta = \pi$. In both the cases when $\theta$ is close to zero, that is all the atoms are initially (at time $t=0$) prepared in their lower states, there develops no quantum correlation among the atoms. This also happens when $\theta =\pi$, that is, initially (at time $t=0$) all the atoms are prepared in their upper states. We observe that correlation increases as the number of atoms in the upper level (in the initial state $|\theta,\phi\rangle$) increases and becomes maximum when the population of the upper level is equal to the population of the lower level, that is, when $\theta = \pi/2$ in the initial state $|\theta,\phi = 0\rangle$. After that, when the population of the excited state increases beyond $50\%$, the correlation
$S$ decreases.
We also see that the net correlation for $N =5$ is higher than that for 
$N =6$ in the lower values of $\theta$. 

In Fig. 5, we show the same plots, but, here the total number of atoms are $N =9$ and $N = 10$. We see that $S$ is asymmetric about $\theta = \pi/2$ for $N =9$, that is an odd number, whereas it is symmetric about $\theta = \pi/2$ for $N = 10$. We also see that, $S$ for $N =9$ is higher than that for $N = 10$, when $\theta$ is small. We observe that, in contrast to Fig. 4, though $S$ increases with $\theta$ but it shows sharp local minimum at 
$\theta = \pi/2$. $S$ goes to zero, when $\theta$ approaches $0$ and 
$\pi$.

In Fig. 6, we show the same graphs with $N = 35$ and $N = 36$. We note that
$S$ for $N = 35$ (odd number) is asymmetric about $\theta = \pi/2$, whereas it is symmetric for $N = 36$. We also note that $S$ is higher for $N = 35$ than that for $N = 36$ in the lower values of $\theta$. Now, both the curves show maxima at $\theta = \pi/2$ and the fluctuations in $S$ near 
$\theta =\pi/2$ dissapears in contrast to Fig. 4 and 5. We also note that 
$S$ goes to $0$ when $\theta$ approaches $0$ and 
$\pi$. 

In Fig. 5 and 6 also, we observe that correlation increases with the increase in the population of the excited state, till the excited state has 
$50\%$ of the total number of atoms, and, after that the correlation decreases gradually when the population of the excited state increases beyond $50\%$.

In Fig. 7, we show the variation of the amount of quantum correlation ($S$) with the interaction time $\tau$, not only for the atomic cat state, but also for the general quantum state of 
$N$ two-level atoms in a dispersive cavity, whose time evolution is governed by the Hamiltonian, given in Eq. (\ref{3.3}), and when the initial state is the atomic coherent state. That is, the value of $\tau$ here is not restricted to the values $\tau = \pi/m$ (where $m$ is an integer), but $\tau$ has been allowed to vary continuously from $0$ to $7$. Thus, the graph shows the variation of $S$ with $\tau$ both for the atomic cat state (when $\tau = \pi/m$) and the general state of 
$N$ two-level atoms in a dispersive cavity. Here, we have taken the total number of atoms $N = 10$, and the value of $\theta$ is $\pi/4$ , that is, the number of atoms in the excited state is one-fourth of the total number of atoms and $\phi$ has been taken as 0. The graph shows that, $S$ increases with $\tau$, attains a maximum, showing oscilattions around the maximum, and then decreases to zero with further increase in $\tau$. This behaviour of $S$ is repeated periodically with $\tau$.

\section{IV. SUMMARY AND CONCLUSION}

We propose a total measure of the multi-particle quantum correlations in a system of $N$ two-level atoms. We take into account all the correlations, that  a system of atoms can possess, that is, the correlations along all the three mutually orthogonal directions and, thus, construct the total measure of quantum correlation of the system of atoms. 
We express the quantum fluctuations $\Delta J_{x^\prime}^2$, 
$\Delta J_{y^\prime}^2$ and $\Delta J_{z^\prime}^2$, of the composite system of atoms, separately,
as an algebraic sum of the corresponding quantum fluctuations of the $N$ individual atoms and the correlation terms, which are made up of all possible combinations of quantum correlations among the atoms. Our ability to write the composite quantum fluctuations in the above way gives us the opportunity to extract out only the quantum correlation factors of the atoms and construct the total measure of the quantum correlation of the system. 
Since, the correlation terms may be positive or negative, we cannot simply add them to have a total measure of the net correlation and, therefore, we take the root mean square value of those correlation terms and define it to be a total measure of the quantum correlation of the whole system of atoms. We use that measure
to quantify the quantum correlation in atomic Schr\"{o}dinger cat states. Schr\"{o}dinger cat states play important role in the conceptual foundations of quantum mechanics. The atomic 
Sch\"{o}dinger cat states, which were proposed by Agarwal et al. \cite{Agarwal1}, are generated using the dispersive interactions in a cavity. These states shows interesting interferences in its quasidisdributions and are highly correlated states. This is our motivation to quantify the amount of quantum correlations present in such states.     

We quantify the net correlation for different atomic cat states corresponding to different superposition of atomic coherent states.  
We study, how the net quantum correlation varies as we increase the number of atoms in the system from $N = 2$ to $N = 100$. We notice one interesting feature, that is, when the number of atoms in the assembly is in between $N =2$ to $N = 10$, the net quantum correlation of odd number of atoms is higher than that for its next higher even number of atoms when the dimensionless interaction time $\tau$ is set to $\pi/6$. After $N =10$, the quantum correlation shows slight zig-zag fashion in its increase with the increase in the number of atoms and finally shows smooth increase when $N$ increases beyond 16.

We also analyse the situation as we increase number of atoms in the upper level in the initial ($t = 0$) state $|\theta,\phi\rangle$ of the atoms. The angle 
$\theta$, here, gives a measure of the number of atoms in the upper state. $\theta = 0$ corresponds to all the atoms in the lower state initially, 
$\theta = \pi/2$ means equal number of atoms in the lower and upper state initially, and $\theta = \pi$ means all the atoms in the upper state initially.  We notice that, the net quantum correlation increases on average as we increase the number of atoms in the excited state. It attains the maximum value when the number of atoms in the upper state is equal to that in the lower state. Then, as we further increase the number of atoms in the upper state, the net quantum correlation starts decreasing and finally goes to zero when all the atoms reach the upper state. But, we also observe that, for low odd number of atoms in the assembly, the net quantum correlation shows irregular pattern in its increase as we increase the number of atoms in the upper state, whereas for even number of atoms, there appears a regular pattern. For odd number of atoms in the assembly, the variation of the net correlation with the angle $\theta$ is asymmetric about 
$\theta = \pi/2$, whereas, for even number of atoms, the corresponding variation is symmetric about $\theta = \pi/2$. We observe small fluctuations in the net correlation near $\theta = \pi/2$, when the number of atoms in the assembly is low, whereas the fluctuations near $\theta = \pi/2$ dissappears when $N = 35$.

We also show the variation of the net quantum correlation with the interaction time $\tau$ for a general state of two-level atoms in a dispersive cavity. We observe that the net correlation increases with the increase in $\tau$, then attains a maximum with small oscillations around the maximum, and then gradually decreases with further increase in $\tau$. This behaviour is repeated periodically with the increase in $\tau$.

Our method can also be applied to the system studied by Felicetti et al., \cite{Felicetti}, where they propose a superconducting circuit architecture for multipartite entanglement generation. It can also be applied to the systems studied in Refs. \cite{Clark} and \cite{Matthias}.

We hope that our study may produce some new insight into the subject.

\end{document}